\newcommand{\chem}[2]{\ensuremath{^{#2}\kern-0.8pt\mathrm{#1}}}
\newcommand{\reac}[6]{\ensuremath{\,^{#2}\kern-0.8pt\mathrm{#1}\,({#3}\,,{#4})\,{}^{#6}\kern-0.8pt\mathrm{#5}\,}}

\newcommand{\Ab}{\ensuremath{\mathcal{A}}}

\newcommand\Gaia{\textit{Gaia}}

\documentclass{aa}  
\usepackage{graphicx}
\usepackage{enumerate}
\usepackage{txfonts}
\usepackage{natbib}

\usepackage{ulem}
\def\modif#1{#1}

\begin{document}
   \title{{\Gaia} eclipsing binary and multiple systems.}

   \subtitle{Two-Gaussian models applied to OGLE-III eclipsing binary light curves in the Large Magellanic Cloud.}

   \author{N.~Mowlavi\inst{1,2}
           \and I.~Lecoeur-Ta\"ibi\inst{2}
           \and B.~Holl\inst{2}
           \and L.~Rimoldini\inst{2}
           \and F.~Barblan \inst{1}
           \and A.~Pr\v sa\inst{3}
           \and A.~Kochoska\inst{4,3}
           \and M.~S\"uveges \inst{5}
           \and L.~Eyer\inst{1}
           \and K.~Nienartowicz\inst{6}
           \and G.~Jevardat\inst{6}
           \and J.~Charnas\inst{2}
           \and L.~Guy\inst{2}
           \and M.~Audard\inst{1,2}
          }

   \institute{Department of Astronomy, Universit\'e de Gen\`eve, 51 chemin des Maillettes, 1290 Versoix, Switzerland\\
              \email{Nami.Mowlavi@unige.ch}
              \and
              Department of Astronomy, Universit\'e de Gen\`eve, 16 chemin d'Ecogia, 1290 Versoix, Switzerland
              \and
              Villanova University, Dept. of Astrophysics and Planetary Science, 800 Lancaster Ave, Villanova PA 19085, USA
              \and
              University of Ljubljana, Dept. of Physics, Jadranska 19, SI-1000 Ljubljana, Slovenia
              \and
              Max Planck Institute for Astronomy, K\"onigstuhl 17, 69117 Heidelberg, Germany
              \and
              SixSq, Rue du Bois-du-Lan 8, CH-1217 Geneva, Switzerland\label{inst3}
             }

   \date{Received ...; accepted ...}

 
  \abstract
   {The advent of large scale multi-epoch surveys raises the need for automated light curve (LC) processing.
    This is particularly true for eclipsing binaries (EBs), which form one of the most populated types of variable objects.
    The {\Gaia} mission, launched at the end of 2013, is expected to detect of the order of few million EBs over a 5-year mission.
   }
   {We present an automated procedure to characterize EBs based on the geometric morphology of their LCs with two aims: first to study an ensemble of EBs on a statistical ground without the need to model the binary system, and second to enable the automated identification of EBs that display atypical LCs.
   }
   {We model the folded LC geometry of EBs using up to two Gaussian functions for the eclipses and a cosine function for any ellipsoidal-like variability that may be present between the eclipses.
   The procedure is applied to the OGLE-III data set of EBs in the Large Magellanic Cloud (LMC) as a proof of concept.
   The bayesian information criterion is used to select the best model among models containing various combinations of those components, as well as to estimate the significance of the components.
   }
   {Based on the two-Gaussian models, EBs with atypical LC geometries are successfully identified in two diagrams, using the Abbe values of the original and residual folded LCs, and the reduced $\chi^2$.
    Cleaning the data set from the atypical cases and further filtering out LCs that contain non-significant eclipse candidates, the ensemble of EBs can be studied on a statistical ground using the two-Gaussian model parameters.
    For illustration purposes, we present the distribution of projected eccentricities as a function of orbital period for the OGLE-III set of EBs in the LMC, as well as the distribution of their primary versus secondary eclipse widths.
   }
   {}

   \maketitle

\section{Introduction}
\label{Sect:introduction}

The interest of binary and multiple systems spans various fields of astrophysics, including stellar formation (initial conditions and formation processes), stellar physics and evolution (accurate stellar parameters determinations and comparison with model predictions), galactic and extra-galactic distance determinations \citep[e.g.][]{Southworth12}, or cosmology (e.g. type Ia supernovae).
Until the end of the twentieth century, binary systems were almost exclusively studied on a case by case basis.
The advent of large scale multi-epoch photometric surveys almost three decades ago with the `Exp\'erience pour la Recherche d'Objets Sombres' \cite[EROS-1, 1990-1995;][]{AubourgBareyreBrehin93,RenaultAubourgBareyre98}, the `Massive Compact Halo Object' experiment \citep[MACHO, 1992-1999;][]{AlcockAllsmanAlves97}, and the `Optical Gravitational Lensing Experiment' \citep[OGLE-I, 1992-1995;][]{UdalskiSzymanskiKaluzny92} opened the door to studies based on large databases containing  thousands to tens of thousands of eclipsing binaries (EBs) in various stellar populations.
Catalogues of EB light curves (LCs) have been published, for example, by the OGLE-III project for the Large Magellanic Cloud \citep[LMC; 26121 sources,][]{GraczykSoszynskiPoleski11}, for the Small Magellanic Cloud \citep[SMC; 6138 sources,][]{PawlakGraczykSoszynski13}, and for the galactic disk fields \citep[11589 sources,][]{PietrukowiczMrozSoszynski13}.
And very recently, the OGLE team updated the list of EBs in the Magellanic Clouds with new results from the OGLE-IV project \citep[40204 sources in the LMC and 8401 sources in the SMC,][]{PawlakSoszynskiUdalski_etal16}.

Another new leap will soon be achieved with on-going and future very large scale multi-epoch surveys that will further increase the number of EBs as well as the level of completeness to an unprecedented degree.
One of those surveys is the European {\Gaia} space mission \citep{PerrymanDeBoerGilmore_etal01,Gaia_PrustiDeBruijneBrown_etal16}, launched in December 2013, the primary aim of which is to determine the three-dimensional positions of over 1 billion stars in the Galaxy.
The preliminary data published in {\Gaia} Data Release 1 \citep{Gaia_BrownVallenariPrusti_etal16} reveals the great potential of the {\Gaia} mission in terms of astrometry, photometry, and number of sources surveyed.
With its combination of all-sky coverage, of multi-epoch white-band photometry (a mean of $\sim$70 photometry transits per source is expected during its 5-year mission), of simultaneous multi-epoch spectro-photometry in blue and red bands, of simultaneous multi-epoch radial velocities and basic astrophysical parameter determinations for the brightest stars, all this in addition to the parallax determinations, the {\Gaia} mission is a golden mine for all fields of astrophysics.
In particular, the mission is expected to record the light curves of between half and several million EBs \citep[e.g.][]{DischlerSoderhjelm05, EyerHollPourbaix_etal13}.
Another example of a promising multi-epoch large scale survey is the photometric Large Synoptic Survey Telescope (LSST) project planned to enter science operations in 2022 \citep{IvezicTysonAcosta_etal08}.

The analysis of hundreds of thousands to millions light (and radial velocity when available) curves from those large scale surveys presents new challenges and requires the development of automated techniques.
Within the {\Gaia} Data Processing and Analysis Consortium, our Geneva-led team is responsible for the detection, characterization and classification of variable objects in general \citep{EyerMowlaviEvans_etal17}, and of EBs in particular.
To achieve these goals on hundreds of thousands of EB LCs, novel processing and analysis techniques are being explored.
The results of those investigations are applied to existing surveys of EBs and simulated {\Gaia} data, and make the object of these series of papers.
Two classification techniques have already been explored, using both existing surveys and {\Gaia} simulated data \citep{SuvegesBarblanLecoeurTaibi_etal17, KochoskaMowlaviPrsa_etal17}.
Here, we present a method to characterize eclipse and inter-eclipse properties based on the geometry of EB folded LCs (FLCs).

The study has two goals.
The first goal is to provide a set of EB parameters that allows to study the ensemble of EBs on a statistical ground without the need to model the binary system.
The second goal is to identify within the large data set binary systems with unexpected properties that could reveal the existence of new configurations.
The procedure is based on modeling the geometry of EB LCs using Gaussian functions to model the eclipses and a cosine function to model ellipsoidal variability, if present.
The models, which we generically refer to in this paper as the two-Gaussian models, whether they actually contain two, one, or no Gaussian, are described in Sect.~\ref{Sect:models}.
The procedure is applied in Sect.~\ref{Sect:OGLEIII} to the set of EBs from the LMC identified by the OGLE-III survey.
The capability of the two-Gaussian models to achieve the two goals is then addressed in Sect.~\ref{Sect:discussion}.
Conclusions are drawn in Sect.~\ref{Sect:conclusions}.

A table summarizing the EB parameters derived in this study for the OGLE-III EBs of the LMC is made available in electronic format.
Its content is described in Appendix~\ref{Appendix:tableDescription}.

\section{Two-Gaussian models}
\label{Sect:models}

We present a description of the geometrical models used to characterize the LCs of EBs (Sect.~\ref{Sect:models_description}), the model computation procedure (Sect.~\ref{Sect:models_computation}), and the best-model selection criterion (Sect.~\ref{Sect:models_selection}).

\subsection{Model description}
\label{Sect:models_description}

\begin{figure}
  \centering
  \includegraphics[width=0.9\columnwidth]{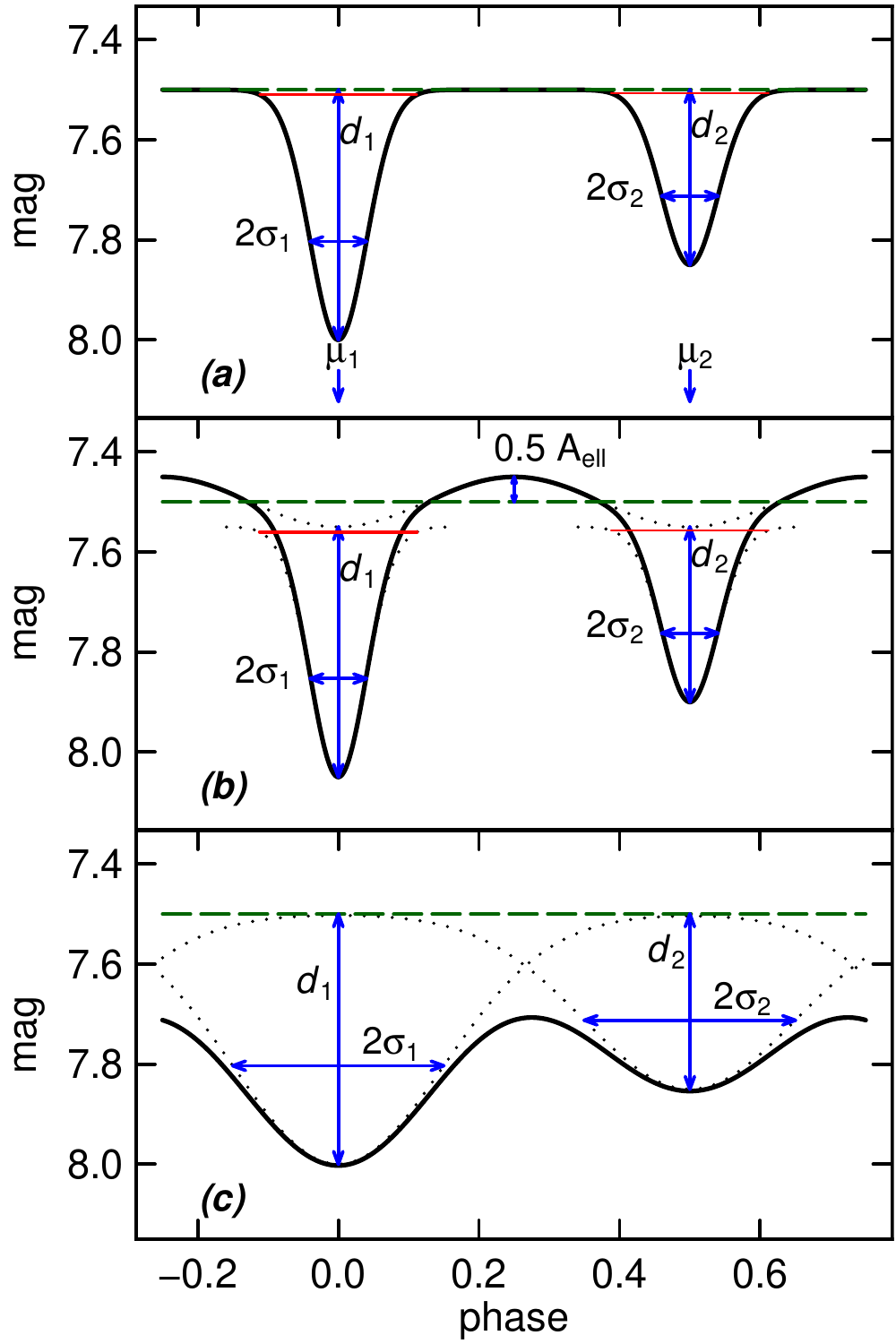}
  \caption{
  Two-Gaussian model parameters used in Eq.~\ref{Eq:gaussianFct} to fit folded light curves of eclipsing binaries.
  The sets of model parameters are, from top to bottom panels:
  \textbf{\textit{(a)}} $C$=7.5~mag, $\mu_1$=0, $d_1$=0.5~mag, $\sigma_1$=0.04, $\mu_2$=0.5, $d_2$=0.35~mag, $\sigma_2$=0.04, $A_\mathrm{Aell}$=0~mag;
  \textbf{\textit{(b)}} same as top panel, but with an ellipsoidal component centered on $\mu_1$ and with $A_\mathrm{ell}$=0.05~mag;
  \textbf{\textit{(c)}} same as top panel, but with $\sigma_1$=0.15 and $\sigma_2$=0.15.
  The green dashed horizontal lines in each panel indicate the value of the constant $C$ in the equation.
  The red continuous horizontal line segments in the top and middle panels give the widths of each of the two Gaussians at 2\% of their depths.
  The black dotted lines in the middle and bottom panels give the individual components of the two-Gaussian models (only the $m=0$ components of the Gaussians in Eq.~\ref{Eq:gaussianFct} are shown).
  The black solid thin lines show the resulting two-Gaussian models.
  }
\label{Fig:gaussianParameters}
\end{figure}

Folded LC geometries are modeled using a Gaussian function for the eclipses, and a cosine function with a period equal to half the orbital period for ellipsoidal-like variability%
\footnote{
The cosine function included in the two-Gaussian model can describe an actual ellipsoidal variability due to tidal interactions, but can also approximate the LC of a semi-detached configuration of a binary system in which one or both stars are partially or fully filling their Roche lobe.
Both effects are referred to, in this paper, as ellipsoidal-like variability.
}%
, if present.

The eclipses are modeled with Gaussian functions of the form
\begin{equation}
  G_{\mu_i,\,d_i,\,\sigma_i}(\varphi) = d_i \; e^{\displaystyle - \frac{(\varphi- {\mu_i})^2}{2\,\sigma_i^2} }\;,
\label{Eq:Gaussian}
\end{equation}
where index $i$ equals 1 and 2 for the primary (deepest) and secondary (least deep) eclipses, respectively, $\mu_i$, $d_i$ and $\sigma_i$ being the Gaussian parameters and $\varphi$ the observation phase (i.e. observation time modulo orbital period).
The ellipsoidal-like variability, on the other hand, is modeled as $\frac{1}{2} \; A_\mathrm{ell} \; \cos [4\pi (\varphi-\varphi_\mathrm{0,ell})]$, where $A_\mathrm{ell}$ is the peak-to-peak amplitude of the ellipsoidal-like variability, and $\varphi_\mathrm{0,ell}$ indicates whether the cosine is centered on eclipse 1 ($\varphi_\mathrm{0,ell}=\mu_1$) or on eclipse 2 ($\varphi_\mathrm{0,ell}=\mu_2)$.
The two-Gaussian model then writes ($C$ is a constant)
\begin{eqnarray}
  G(\varphi) = C & + & \sum_{m=-2}^2 G_{\mu_1+m,\,d_1,\,\sigma_1}(\varphi)
                       \;+\; \sum_{m=-2}^2 G_{\mu_2+m,\,d_2,\,\sigma_2}(\varphi) \nonumber\\
                 & + & \frac{1}{2} \; A_\mathrm{ell} \; \cos [4\pi (\varphi-\varphi_\mathrm{0,ell})] \;,
\label{Eq:gaussianFct}
\end{eqnarray}

Equation~\ref{Eq:gaussianFct} includes the mirrors of eclipses 1 and 2 at phases from -2 to +2 in order to take into account the contribution of the tails of the Gaussian functions from adjacent phases due to the periodicity of the eclipses.
The model parameters are illustrated in Fig.~\ref{Fig:gaussianParameters} for three types of EBs.

By convention, we shift LC times such as to locate the primary eclipse at phase 0.
We therefore always have
\begin{equation}
  \mu_1=0 \; ,
\label{Eq:mu1}
\end{equation}
even though we may continue to explicitly write $\mu_1$ for clarity in some expressions.

Eclipse durations $w_i$ (durations expressed in phase) are taken equal to the widths of the Gaussian functions at a magnitude depth of 2\% relative to Gaussian depth $d_i$, i.e. $w_i = 5.6\,\sigma_i$, with an upper limit of 0.4.
This somewhat arbitrary limit is set in order to avoid unphysical large eclipse durations for wide Gaussians.
We thus have
\begin{equation}
  w_i = \min(\, 5.6\,\sigma_i \;,\; 0.4 \,).
\label{Eq:eclipseWidth}
\end{equation}
Eclipse depths $d'_i$ are taken equal to the difference between the magnitude at the bottom of the eclipse and the brightest magnitude $G_\mathrm{min}$ of the model:
\begin{equation}
  d'_i = G_\mathrm{max}(\mu_i) - G_\mathrm{min}  \;\; .
\label{Eq:eclipseDepth}
\end{equation}
Finally, we note that the constant $C$ in Eq.~\ref{Eq:gaussianFct} equals $G_\mathrm{min}$ only for detached EBs that do not show ellipsoidal-like variability (illustrated in the top panel of Fig.~\ref{Fig:gaussianParameters}).
For EB LCs displaying ellipsoidal-like variability (middle panel of Fig.~\ref{Fig:gaussianParameters}) or for contact binaries (bottom panel of Fig.~\ref{Fig:gaussianParameters}), $C \ne G_\mathrm{min}$.

\subsection{Model computation}
\label{Sect:models_computation}

We fix the orbital period of each EB to the value published in the OGLE-III catalogue.

A two-Gaussian model $G(\varphi)$ defined by Eq.~\ref{Eq:gaussianFct} is fitted to the FLC $\{y_j(\varphi_j)\}$ of each EB, where $j$ is an index running over all measurements from 1 to the number $N_\mathrm{obs}$ of observations.
The computation of the model parameters follows three steps:
time series outlier removal (Sect.~\ref{Sect:procedure_outliers}),
initial values estimation of the two-Gaussian model parameters (Sect.~\ref{Sect:procedure_initialValues}),
and non-linear fitting (Sect.~\ref{Sect:procedure_nonlinearFit}).

\subsubsection{Light curve outliers removal}
\label{Sect:procedure_outliers}

Outlier removal is performed in two steps.
First, all measurements with uncertainties greater than 1 magnitude are removed.
Second, isolated measurements having magnitudes at the extremes of the magnitude distribution are removed.
To do this, measurements with extreme magnitudes are identified from their deviations from the median magnitude when these exceed a certain number of times the inter-quantile range IQR (10 times at the faint side and 2 times at the bright side).
They are considered to be outliers, and removed from the time series, unless they have similar (magnitude within 30\%) neighbors in time (preceding or following measurement in time within 1/4 days) or in the magnitude distribution (nearest points in the histogram of magnitudes).



\subsubsection{Initial value determination of model parameters}
\label{Sect:procedure_initialValues}

Fitting a two-Gaussian model to a time series is very sensitive to the adopted initial values of the parameters.
The better the initialization is, the better the convergence of the non-linear fitting algorithm is expected.
We therefore proceed in three steps, first to catch the global shape of the FLC, then to detect the two eclipse candidates, and finally to initialize the two-Gaussian model.

\paragraph{Folded light curve smoothing}

We start performing a weighted running average on the FLC, replacing each magnitude value $y_j$ at a given phase $\varphi_j$ by a weighted average $\tilde{y}_j$ of the magnitudes $y_k$ within a $[\varphi_j-\delta \varphi, \varphi_j+\delta \varphi]$ phase window.
The weights $w_k$ are taken equal to
\begin{equation}
  w_k = e^{- \frac{(\varphi_k - \varphi_j)^2}{2 \; \delta \varphi^2}} ,
\end{equation}
and the average magnitude is given by
\begin{equation}
 \tilde{y}_j = \frac{\sum_k {w_k \; y_k}}{\sum_k w_k}
\end{equation}
with the index $k$ running on all measurements available in the phase window.
We take $\delta \varphi = 0.01$.
From this FLC \{$\tilde{y}_j$\}, an evenly sampled FLC of 200 points is produced by linear interpolation of the averaged FLC.
A smoothed FLC is then computed using the Savitzky-Golay (SG) algorithm \citep{SavitzkyGolay64,Gorry90,ProtopapasGiammarcoFaccioli_etal06}, which has the main advantage to preserve quite well the minima and widths of the eclipses.
We use the Java implementation of the SG algorithm in the Flanagan library\footnote{\url{www.ee.ucl.ac.uk/~mflanaga}}, which consists of a least-squares polynomial regression of degree 3 applied on $2M+1$ points centered on each considered point, $M$ being a parameter which we take equal to 15.
The resulting smooth FLC is denoted the SG FLC.

\paragraph{Eclipse identification}

Eclipse candidates are searched for in the SG FLC.
We define a threshold magnitude equal to the median magnitude plus the median of the observation uncertainties, and determine a baseline magnitude $M_b$ equal to the median magnitude of all observations brighter than this threshold.
We then select the two faintest dips having magnitudes above this baseline as the two eclipse candidates.

\paragraph{Initial value estimation of model parameters}

The initial value of the constant $C$ is set to the baseline magnitude $M_b$ computed in the preceding step.
The initial value of $\mu_1$ ($\mu_2$) is set equal to the phase of the measurement closest to the maximum magnitude of the deepest (second deepest) eclipse candidate identified in the SG FLC, while $d_1$  ($d_2$) is set to the difference between that maximum magnitude and the baseline magnitude.
Finally, $\sigma_1$ ($\sigma_2$) is taken equal to 0.2 times the phase extent covered by all adjacent measurements around $\mu_1$ ($\mu_2$) fainter than the baseline magnitude. 

\subsubsection{Non-linear fitting procedure}
\label{Sect:procedure_nonlinearFit}

We use the non-linear fitting algorithm \texttt{nls} of the R Project for Statistical Computing to search the solution to Eq.~\ref{Eq:gaussianFct}.
We estimate the initial values of the parameters as explained in the previous section, swapping the order of the eclipses if necessary to have the first eclipse to be the deepest.
\modif{A weight of $1/\varepsilon_i^2$ is assigned to each measurement $y_i$, where $\varepsilon_i$ is the uncertainty on $y_i$.}
We constrain the solutions to positive Gaussian depths and cosine amplitudes (the later constrain to avoid the non-linear method to converge to a sine solution) by transforming Eq.~\ref{Eq:gaussianFct} such that it takes the logarithm of $d_1$, $d_2$ and $A_\mathrm{ell}$.
If, after convergence, the second eclipse turns out to be deeper than the first eclipse, we swap the two eclipses and search again for a solution because the cosine variability, if present, may impact the solution when the two eclipse candidates are not separated by exactly 0.5 in phase.
This procedure ensures a consistent solution with the primary eclipse always numbered 1.

\subsection{Model selection}
\label{Sect:models_selection}

\begin{table}
\centering 
\caption{Two-Gaussian models used to describe eclipsing binary light curve geometries.
         The last column gives the number of parameters in the models.
} 
\begin{tabular}{l l c} 
\hline 
Model      & Description                        & Nbr of \\ 
           &                                    & params \\ 
\hline \\[-2mm] 
\multicolumn{2}{l}{\textit{--- Two eclipses}}   &            \\[1mm] 
CG12       & Without ellipsoidal-like var.      &          7 \\ 
CG12E1     & With ellipsoidal-like var. on eclipse 1 &     8 \\ 
CG12E2     & With ellipsoidal-like var. on eclipse 2 &     8 \\ 
\hline \\[-2mm] 
\multicolumn{2}{l}{\textit{--- One eclipse}}    &            \\[1mm] 
CG         & Without ellipsoidal-like var.      &          4 \\ 
CGE        & With ellipsoidal-like var. on eclipse 1 &     5 \\ 
\hline \\[-2mm] 
\multicolumn{2}{l}{\textit{--- No eclipse}}   &            \\[1mm] 
CE         & Ellipsoidal-like var.              &          3 \\ 
C          & Constant                           &          1 \\ 
\hline 
\end{tabular}
\label{Tab:models}
\end{table}

Several models are tested on each LC, and their Bayesian information criterion (BIC) compared to identify the model that best matches the data given the measurement uncertainties.
The BIC is computed as \citep[][Eq.~3.54]{FeigelsonBabu12}
\begin{equation}
  \mathrm{BIC} = 2 * \ln L - p * \ln N_\mathrm{obs}
\label{Eq:BIC}
\end{equation}
where $p$ is the number of model parameters, given in Table~\ref{Tab:models} for the models considered in this paper, and $\ln L$ is the log-likelihood given by
\begin{equation}
  \ln L = - \sum_{j=1}^{N_\mathrm{obs}} \left\{ \ln \left(\sqrt{2\pi} \; \varepsilon_j \right)
                 + \frac{[y_j - G(\varphi_j)]^2}{2\;\varepsilon_j^2} \right\}
\end{equation}
with $\varepsilon_j$ being the uncertainty on measurement $y_j$.

Including two Gaussian functions and a cosine in Eq.~\ref{Eq:gaussianFct} to model FLCs may lead to an overfit of the data if one or more of the components are insignificant (e.g., if the amplitude of the given component is small compared to the mean uncertainty of the measurements) or spurious (e.g., if the locations of the eclipse candidates were wrongly initialized).
We therefore fit several models to each FLC, each having a different combination of components, and retain the one that has the highest BIC.
The BIC takes into account the number of degrees of freedom of each model.
We therefore avoid \modif{overfitting} the data \modif{with} non-significant components (either Gaussians or ellipsoidal variability).

The various models are summarized in Table~\ref{Tab:models}.
They comprise:
\begin{itemize}
\item a pure constant model (model 'C'), representing a LC with no detectable eclipse or ellipsoidal-like variability;
\vskip 2mm
\item a model including only an ellipsoidal component (model 'CE');
\vskip 2mm
\item models including only one Gaussian, without (model 'CG') or with (model 'CGE') an ellipsoidal component.
For the CGE model, the $\varphi_\mathrm{0,ell}$ parameter in Eq.~\ref{Eq:gaussianFct} is taken equal to the $\mu$ value of the eclipse candidate;
\vskip 2mm
\item models including two Gaussians, without (model 'CG12') or with (models 'CG12E1' and 'CG12E2') an ellipsoidal component.
Models 'CG12E1' and 'CG12E2' distinguish cases where the cosine of the ellipsoidal component is centered on the first ($\varphi_\mathrm{0,ell} = \mu_1$) or second ($\varphi_\mathrm{0,ell} = \mu_2$) eclipse candidate, respectively.
They differ from one another only for eccentric systems for which $\mu_2-\mu_1 \ne 0.5$.
\end{itemize}
The initial values of the model parameters for each of those models are taken from the set of initial values computed in Sect.~\ref{Sect:procedure_initialValues}.
When testing models with only one Gaussian, if the procedure described in Sect.~\ref{Sect:procedure_initialValues} identifies two eclipse candidates, two sets \{CG, CGE\} of models are tested, one set \{CG1, CG1E1\} with the Gaussian (and cosine when present) centered on the first eclipse candidate and another set \{CG2, CG2E2\} with the Gaussian (and cosine when present) centered on the second eclipse candidate.
Those two sets are not distinguished in Table~\ref{Tab:models}, where models CG1 and CG2 (or CG1E1 and CG2E2) are indistinguishably referred to as model CG (or CGE).

Having computed, for a given FLC, all the models mentioned above, their BIC values are compared with one another, and the model with the highest BIC is retained.
One exception to this rule concerns models with ellipsoidal-like variability that contain a wide Gaussian.
Those models are retained for comparison with the other models only if the phase duration of the eclipse(s) is(are) shorter than a given limit $w_\mathrm{max,ell}$ after model convergence, i.e. if $\sigma < w_\mathrm{max,ell}/5.6$ for models with one Gaussian, or if both $\sigma_1 < w_\mathrm{max,ell}/5.6$ and $\sigma_2 < w_\mathrm{max,ell}/5.6$ for models having two Gaussians 
(we take $w_\mathrm{max,ell}=0.4$).
Otherwise, the model is automatically rejected in favor of the models without ellipsoidal component.
This condition is imposed in order to lift the degeneracy of some \textsl{EB}-type binaries for which the FLC can be modeled, for example, by either a CG12 model (correct model) or by a CGE model that includes a wide Gaussian on top of an ellipsoidal-like variability (fake model where the wide Gaussian added to one of the two depths of the ellipsoidal-like variability mimics a two-eclipse \textsl{EB}-type binary with non-equal depths).
Tests performed on some \textsl{EB}-type binaries showed that the BIC value of the CGE model could indeed be larger than that of the CG12 value, thereby leading to a wrong model selection.
Similarly to this example for models with one Gaussian, models with two Gaussians and an ellipsoidal component (models CG12E1 and CG12E2) are a-posteriori discarded if the phase duration of either Gaussian is larger than $w_\mathrm{max,ell}$.

\section{Application to OGLE-III LMC eclipsing binaries}
\label{Sect:OGLEIII}

We apply in this section the two-Gaussian model to the \textit{I}-band LCs of the 26121 EBs in the LMC published by the OGLE-III survey \citep{GraczykSoszynskiPoleski11}.
The survey operated from July 2001 to May 2009.
Each LC has, in the mean, 500 measurements, 90\% of which are in the photometric \textit{I} band.
The LCs and original data are all downloaded from \texttt{ftp://ftp.astrouw.edu.pl/ogle3/}.

We fix the orbital periods to the values listed by \cite{GraczykSoszynskiPoleski11}.
The LCs are then fitted by a two-Gaussian model following the procedure described in Sect.~\ref{Sect:models}, selecting the best model among a combination of Gaussian and cosine functions based on the BIC analysis (see Sect.~\ref{Sect:models_selection}).
\modif{The computation takes less than 1 sec per source on a single 2.7~GHz CPU.}

The vast majority (85\%) of EB LCs are modeled with two Gaussians (with or without an additional ellipsoidal component), and 14\% of LCs are modeled with only one Gaussian (with or without an ellipsoidal component).
The remaining 1\% EBs have their LCs modeled with only an ellipsoidal component, except for \modif{two} cases \modif{for which the highest BIC model is a pure constant}.

The success of our procedure to model the FLCs of the majority of OGLE-III EBs with two Gaussians does not guarantee reliable model components.
The efficiency to identify true components in the LCs can be hampered by several effects, including time sampling, measurement uncertainties, wrong initial guess of eclipse locations, or additional intrinsic variability in one or both stars of the binary system.
Some Gaussian or ellipsoidal-like components in the final models may therefore be missing, spurious, or wrong.
We therefore devote this section to analyse model results.

We first present in Sect.~\ref{Sect:phaseCoverage} the phase coverage properties of the OGLE-III LMC EBs.
Phase coverage depends only on observation time sampling and orbital period, but can impact model results.
The significance and reliability of model components are then studied in Sect.~\ref{Sect:componentsSignificance}.
The question of model degeneracy is addressed in Sect.~\ref{Sect:modelsDegeneracies}, and the quality of models is analyzed in Sect.~\ref{Sect:fitQuality}.
Finally, Sect.~\ref{Sect:tableSummary} mentions the table published in electronic format that provides the results of this paper for the OGLE-III LMC EBs.

\subsection{Phase coverage}
\label{Sect:phaseCoverage}

\begin{figure}
  \centering
  \includegraphics[width=1.00\columnwidth]{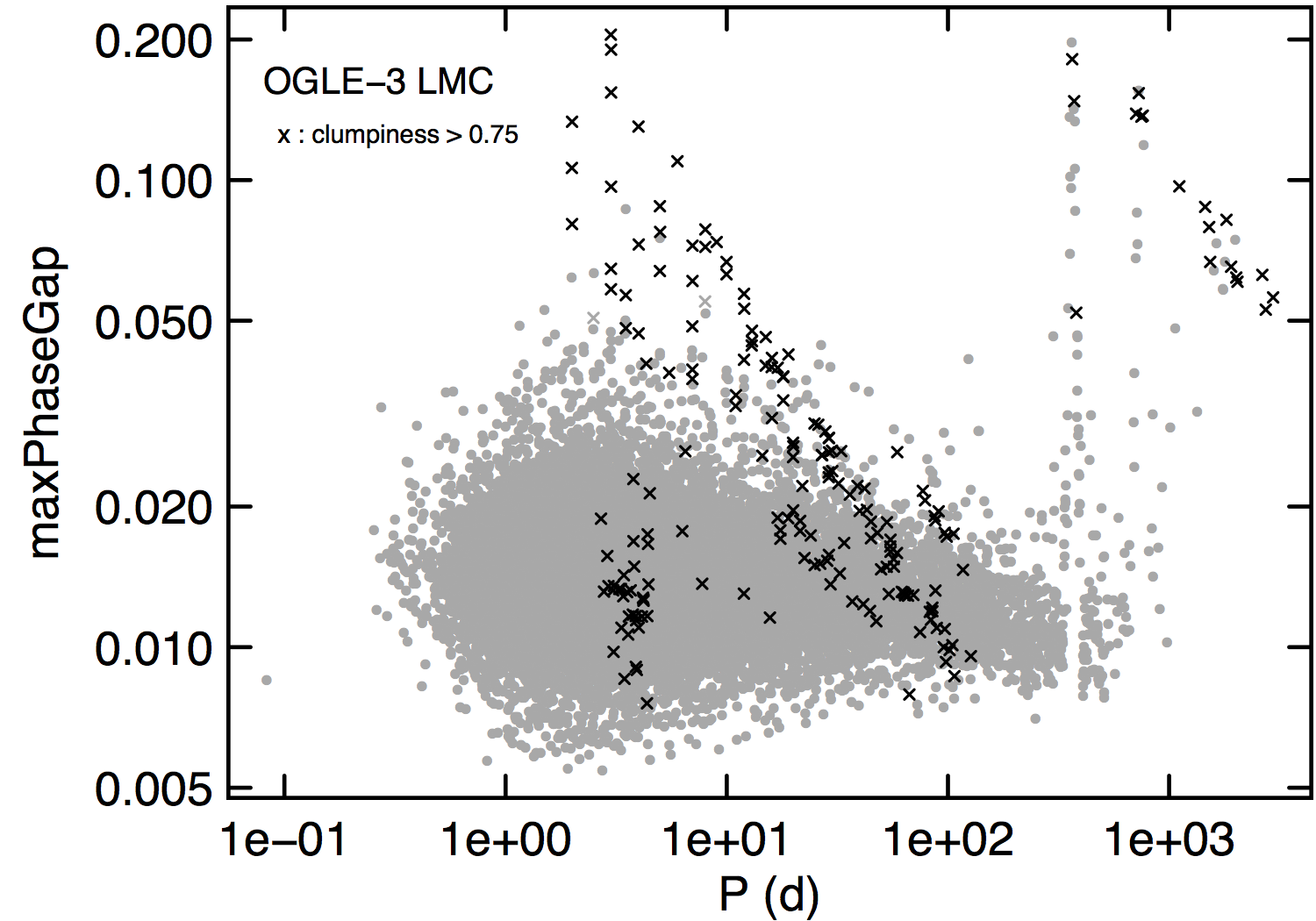}
  \caption{Maximum phase gap versus period for all OGLE-III LMC eclipsing binaries.
           Folded light curves with a clumped distribution of their phases (clumpiness above 0.75 on a scale between 0.5 for a uniform-like distribution to 1 for a highly clumped distribution) are shown as black crosses, while the other folded light curves (clumsiness below 0.75) are shown as gray filled circles.
           }
\label{Fig:maxPhaseGap}
\end{figure}

\begin{figure}
  \centering
  \includegraphics[width=1.00\columnwidth]{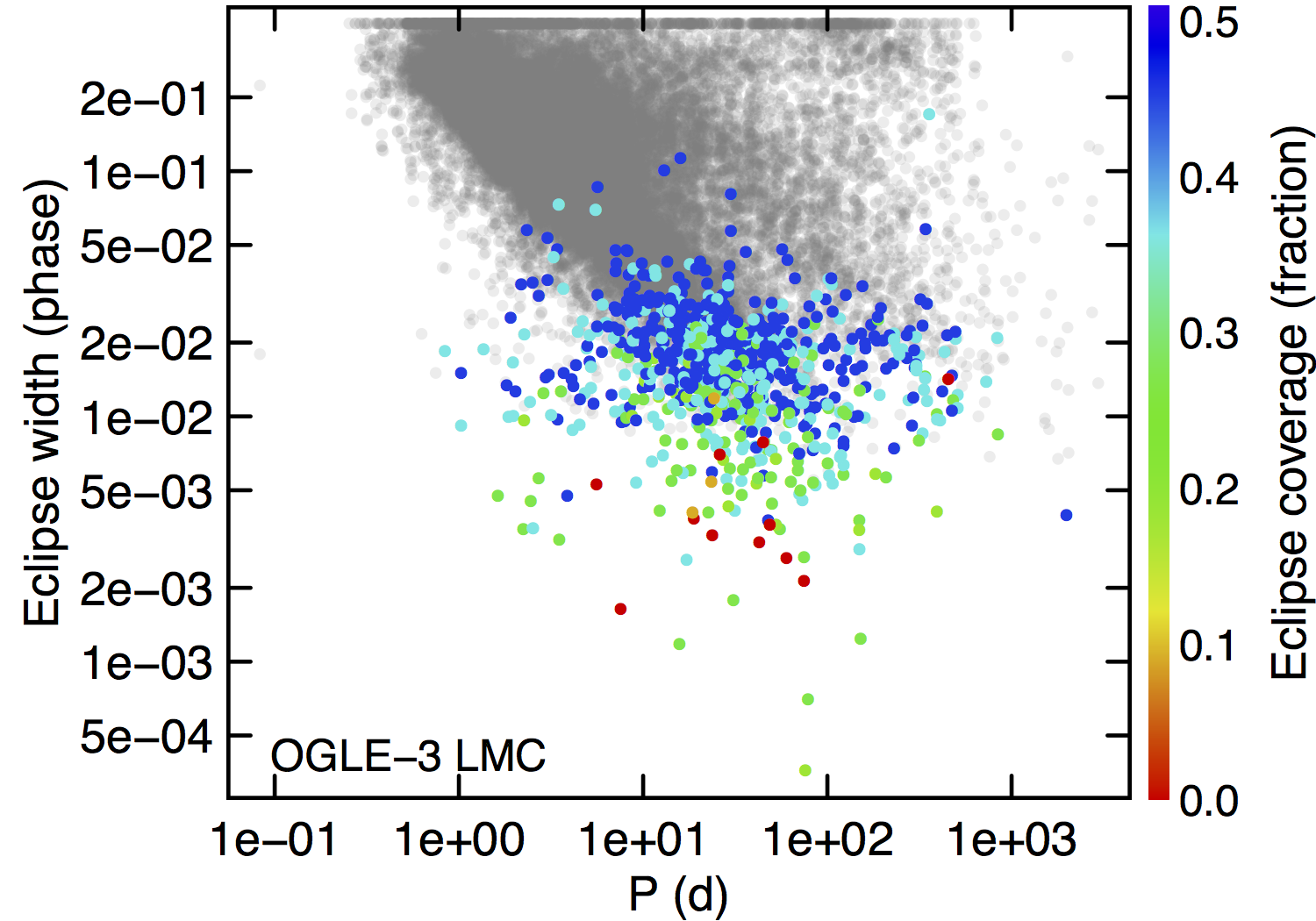}
  \caption{Eclipse phase width (Eq.~\ref{Eq:eclipseWidth}) versus orbital period of the eclipsing binaries.
           A color is used if the observations cover less than 50\% of the eclipse width, with the color indicating the phase coverage fraction according to the color scale on the right of the figure.
          }
\label{Fig:eclipsePhaseCoverage}
\end{figure}

A good phase coverage of the eclipses is essential to correctly model the LCs.
This, however, depends on the observation time sampling, on eclipse duration and on orbital period.

For a ground-based survey like OGLE, a significant phase gap is expected for orbital periods close to a multiple of the day (since the star is observable only during nights) or of the year (since the star is visible only at specific times of the year).
The phases are then clumped in groups.
Figure~\ref{Fig:maxPhaseGap} plots the largest phase gap recorded in each OGLE-III LMC EBs FLCs versus their orbital period, highlighting in black crosses the FLCs that have phase-clumped data.
The degree of clumpiness is evaluated based on the distribution of phase intervals between two successive measurements in the FLC.
We do this after having shifted the phases of the FLCs by a constant value such as to move the largest phase interval to the end of the [0,1] phase interval (this is done in order to have a quantity that is independent of the reference time used to fold the time series).
We then define the phase clumpiness as the fraction of phase intervals that have durations less than $1/(N_\mathrm{obs}-1)$.
The phase clumpiness computed in this way is expected to be around 0.5 for sources regularly sampled in phase, and close to 1 for highly-clumped distributions in phase.
Sources with a phase clumpiness above 0.75 are highlighted with black crosses in Fig.~\ref{Fig:maxPhaseGap}.
Their two-Gaussian model parameters may be at fault due to missing data.

Another useful phase coverage related quantity is eclipse phase coverage by observations.
We estimate the eclipse coverage by considering 11 equal phase intervals within the eclipse width $[\mu_i-w_i, \mu_i+w_i]$ and by computing the fraction of these intervals that have at least one measurement.
A value of 0 would mean that no observation is available within the considered interval --~this can happen if observations are only available at the very borders of the eclipse candidate~--, while a value of 1 means that measurements are available in all eleven phase bins.
Eclipse candidates with insufficient eclipse coverage may have wrong model parameters.
Such eclipse candidates are usually, but not always, spurious.
About \modif{92}\% of the sources in our sample have their eclipse candidates covered by observations over more than 70\% of their durations.
Eclipses with less than a few percent coverage are usually narrow in phase, irrespective of their period.
This is shown in the eclipse width versus orbital period diagram displayed in Fig.~\ref{Fig:eclipsePhaseCoverage}, where eclipses with a  phase coverage of less than 50\% are highlighted in color.

\subsection{Model components significance}
\label{Sect:componentsSignificance}

The reliability of the Gaussian and ellipsoidal components found by the two-Gaussian model procedure is analyzed in this section.
The analysis is done in Sect.~\ref{Sect:modelsSignificance_BIC} based on the BIC values obtained for different models.
The significance of the eclipses and of the ellipsoidal component are then considered in Sects.~\ref{Sect:modelsSignificance_eclDepth} and \ref{Sect:modelsSignificance_ellAmplitude}, respectively.

\subsubsection{Reliability of two-Gaussian model components}
\label{Sect:modelsSignificance_BIC}

\begin{figure}
  \centering
  \includegraphics[width=1.00\columnwidth]{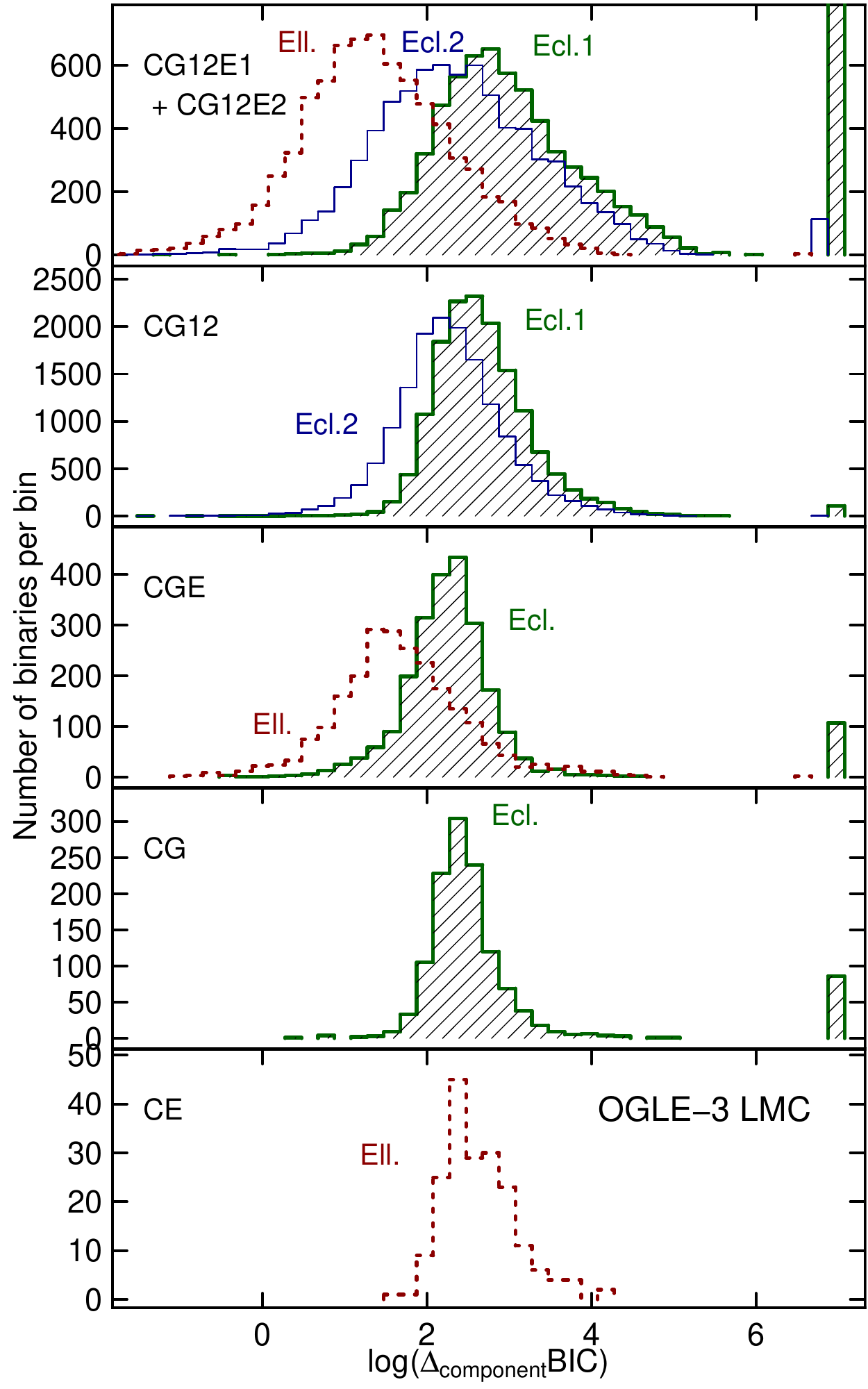}
  \caption{Distributions of the BIC value differences between the best model chosen by the automated two-Gaussian procedure and the alternative model without eclipse 1 (thick green hatched histogram), the alternative model without eclipse 2 (thin blue histogram), and the alternative model without the ellipsoidal component (dashed red histogram).
          The distributions are shown for best models that contain two Gaussians and an ellipsoidal-like component (top panel), two Gaussians only (second panel from top), one Gaussian and an ellipsoidal-like component (third panel from top), one Gaussian only (fourth panel from top), and an ellipsoidal-like component only (bottom panel).
           The histograms are plotted as a function of the logarithm (base 10) of the BIC value differences, with a bin width of 0.2.
          The number of models for which the alternative model did not converge or had a negative infinite BIC value is shown on the right of each panel at an x-axis value of 7 (for the eclipse 1 component), 6.8 (for the eclipse 2 component) and  6.6 (for the ellipsoidal component).
          In the top panel, the Y-axis is limited to \modif{700} for a better visibility, \modif{1299} models having no alternative model without eclipse 1.
           }
\label{Fig:histoComponentsReliability}
\end{figure}

\begin{figure}
  \centering
  \includegraphics[width=1.00\columnwidth]{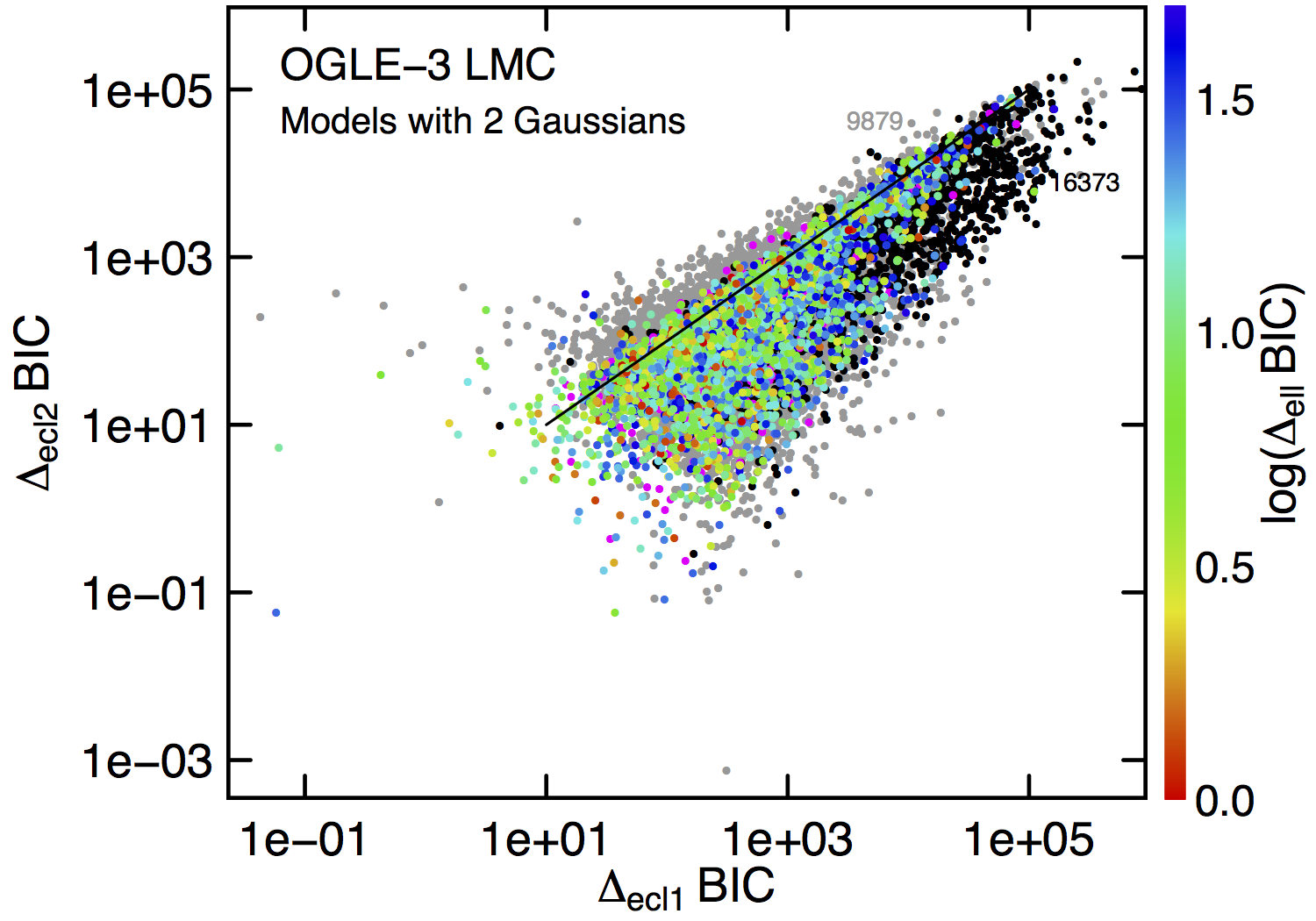}
  \caption{Secondary eclipse significance $\Delta_\mathrm{ecl2} \mathrm{BIC}$ versus primary eclipse significance $\Delta_\mathrm{ecl1} \mathrm{BIC}$ of all models that contain two Gaussians.
           Models without ellipsoidal component are plotted in gray.
           Models containing an ellipsoidal component are shown in color, the color being related to $\Delta_\mathrm{ell} \mathrm{BIC}$ according to the color scale drawn on the right of the figure.
           Models that have $\log(\Delta_\mathrm{ell} \mathrm{BIC})$ values greater (smaller) than the upper (lower) limit shown on the color scale are plotted in black (magenta).
           A 1:1 line is added to the figure as an eye-guide.
           The sources labeled in the figure have their folded light curves displayed in Fig.~\ref{Fig:flcsSignificantComponents}.
           }
\label{Fig:BicDiffComparison_twoGaussians}
\end{figure}

\begin{figure}
  \centering
  \includegraphics[width=0.95\columnwidth]{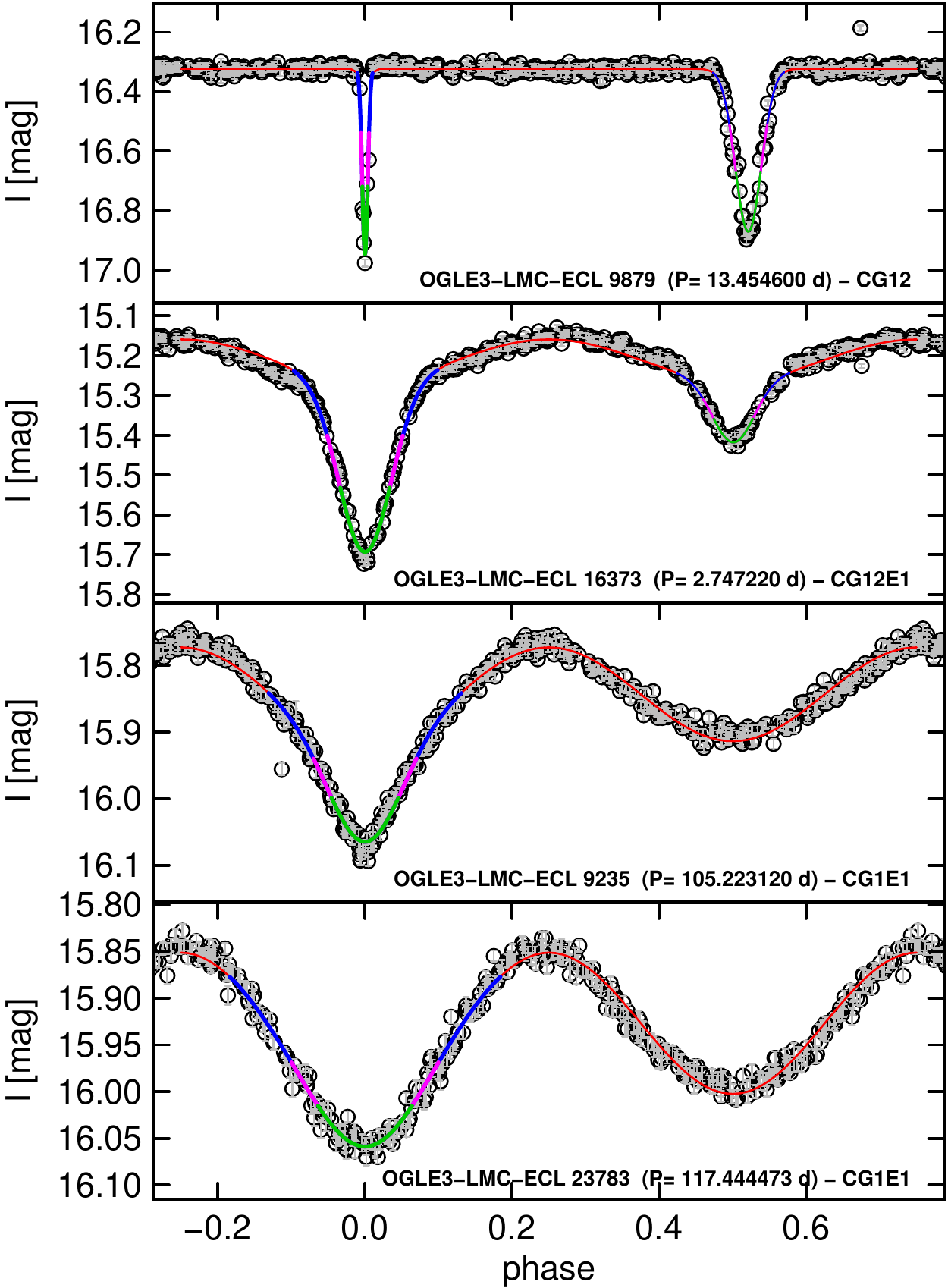}
  \caption{Example folded light curves with significant eclipse and ellipsoidal components\modif{, with the model with the highest BIC indicated in each panel}.
           The models include two (one) Gaussians for the sources shown in the two top (two bottom) panels.
           \modif{The green, magenta and blue segments of the model show the eclipse extensions up to $\mu_i\pm\sigma_i$, $\mu_i\pm1.5\sigma_i$, and $\mu_i\pm2.8\sigma_i$, respectively.
           The red parts of the model indicate out-of-eclipse region (based on an eclipse phase width of $5.6 \sigma$}.
           \modif{If the Gaussians have $\sigma>0.5/5.6$, the whole model is drawn in red (this is the case for some sources in other example light curves in this paper).}
           The sources are labeled in Fig.~\ref{Fig:BicDiffComparison_twoGaussians} (Fig.~\ref{Fig:BicDiffComparison_oneGaussian}).
           }
\label{Fig:flcsSignificantComponents}
\end{figure}

\begin{figure}
  \centering
  \includegraphics[width=1.00\columnwidth]{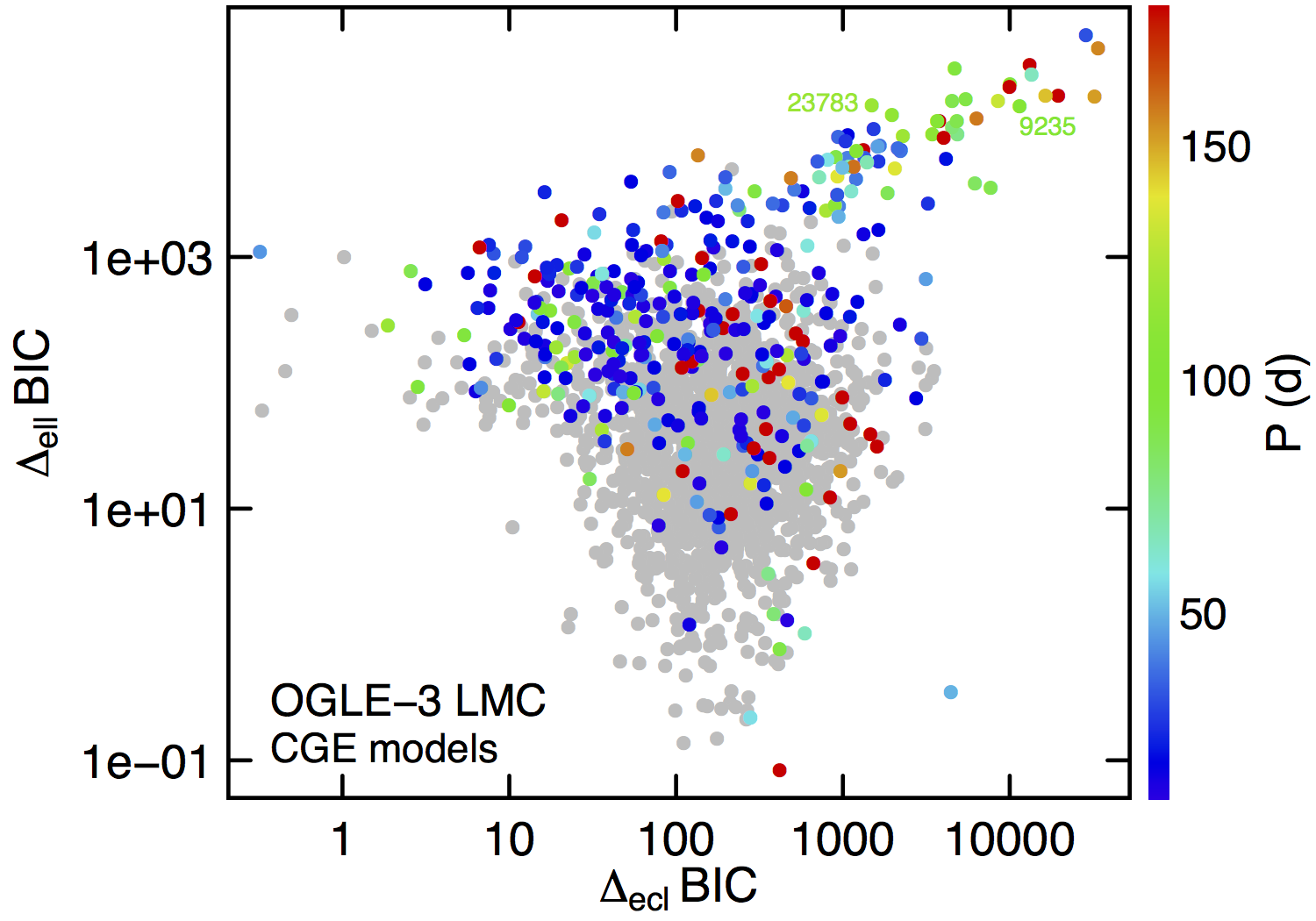}
  \caption{Ellipsoidal component significance $\Delta_\mathrm{ell} \mathrm{BIC}$ versus eclipse significance $\Delta_\mathrm{ecl} \mathrm{BIC}$ of models having one Gaussian and an ellipsoidal component.
           The color is related to orbital period according to the color scale drawn at the right of the figure, with orbital periods smaller than 10~days plotted in gray and those larger than 180~days plotted in red.
           The sources labeled in the figure have their folded light curves displayed in Fig.~\ref{Fig:flcsSignificantComponents}.
           }
\label{Fig:BicDiffComparison_oneGaussian}
\end{figure}

The reliability of a model component detected by the two-Gaussian procedure can be estimated by comparing the BIC value of the given model to the BIC value of the alternative model without the given component.
The difference between these two BIC values, BIC(with component) $-$ BIC(without component), is denoted by $\Delta_\mathrm{component} \mathrm{BIC}$.
The distribution of $\Delta_\mathrm{Ecl1} \mathrm{BIC}$ for the primary eclipse candidate, of $\Delta_\mathrm{Ecl2} \mathrm{BIC}$ for the secondary eclipse candidate, and of $\Delta_\mathrm{Ell} \mathrm{BIC}$ for the ellipsoidal component are shown in Fig.~\ref{Fig:histoComponentsReliability} for the various models (from models containing two Gaussians and an ellipsoidal component in the top panel to models containing only an ellipsoidal component in the bottom panel).
The larger the $\Delta_\mathrm{component} \mathrm{BIC}$ difference is, the larger the probability is for the given component to be significant and non-spurious. In the great majority of cases, the reliability of the eclipse candidates is very good, with \modif{91}\% (\modif{66}\%) of models with two Gaussians satisfying $\Delta_\mathrm{ecl1} \mathrm{BIC}>100$ ($\Delta_\mathrm{ecl2} \mathrm{BIC}>100$) for the primary (secondary) eclipse irrespective of the presence of an ellipsoidal component, and \modif{81}\% of models with one Gaussian satisfying $\Delta_\mathrm{ecl} \mathrm{BIC}>100$ irrespective of the presence of an ellipsoidal component.

The significances of the various components in models containing two Gaussians are shown in Fig.~\ref{Fig:BicDiffComparison_twoGaussians}, where $\Delta_\mathrm{ecl2} \mathrm{BIC}$ is plotted versus $\Delta_\mathrm{ecl1} \mathrm{BIC}$ with $\Delta_\mathrm{Ell} \mathrm{BIC}$ shown in color for models with an ellipsoidal component.
Two examples with highly significant eclipses are shown in the top panels of Fig.~\ref{Fig:flcsSignificantComponents}.
Small values of $\Delta_\mathrm{ecl1} \mathrm{BIC}$ or $\Delta_\mathrm{ecl2} \mathrm{BIC}$, on the other hand, point to unreliable primary or secondary eclipse candidates, respectively.
Fortunately, this concerns only a small fraction of the eclipse candidates, as seen in Fig.~\ref{Fig:BicDiffComparison_twoGaussians}.
This feature must however be kept in mind when studying the ensemble of EBs with the two-Gaussian model results.

A similar analysis can be done on models containing one Gaussian.
The distribution of $\Delta_\mathrm{ecl} \mathrm{BIC}$ and $\Delta_\mathrm{ell} \mathrm{BIC}$ for the ones containing an ellipsoidal component is displayed in Fig.~\ref{Fig:BicDiffComparison_oneGaussian}, and two examples with highly significant components are shown in the two bottom panels of Fig.~\ref{Fig:flcsSignificantComponents}.

\subsubsection{Eclipse significance}
\label{Sect:modelsSignificance_eclDepth}

\begin{figure}
  \centering
  \includegraphics[width=1.00\columnwidth]{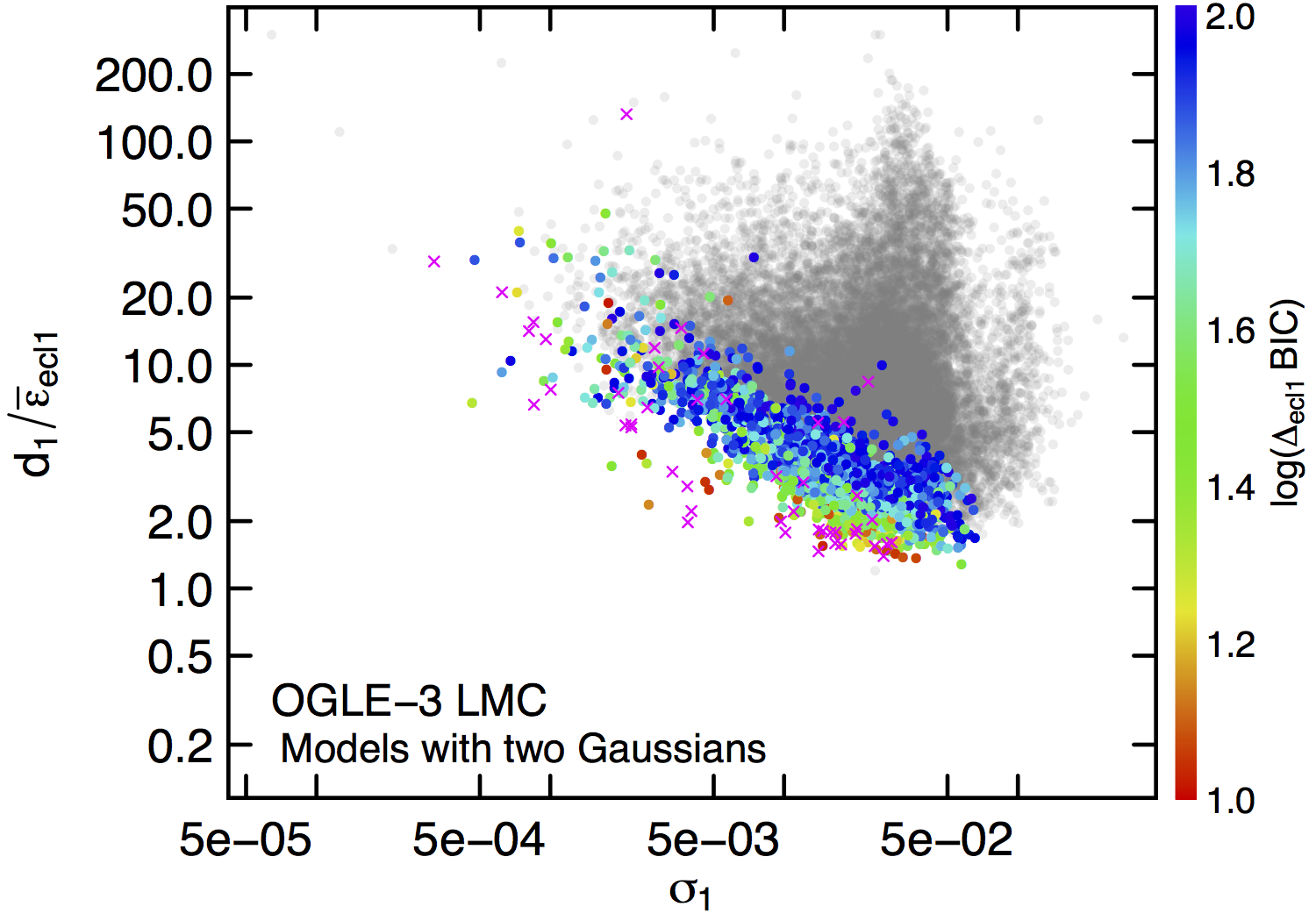}
  \caption{Ratio of Gaussian depth $d_1$ of primary eclipse candidates to mean measurement uncertainty $\bar{\varepsilon}_{\mathrm{ecl},1}$ inside the eclipse, versus Gaussian widths $\sigma_1$ of all models containing two Gaussians, with or without an ellipsoidal component.
           For a better visibility, all $d_1/\bar{\varepsilon}_{\mathrm{ecl},1}$ ratios larger than 300 are plotted on the Y-axis at the value of 300.
           The color of the markers is related to the $\Delta_\mathrm{ecl1} \mathrm{BIC}$ differences between the BIC of the adopted model and the BIC of the corresponding model without the primary eclipse.
           A gray color is used for BIC differences larger than 50, and a magenta color for BIC differences smaller than 10.
           }
\label{Fig:dOverMeanMagErrorVsSigma_ecl1}
\end{figure}

\begin{figure}
  \centering
  \includegraphics[width=1.00\columnwidth]{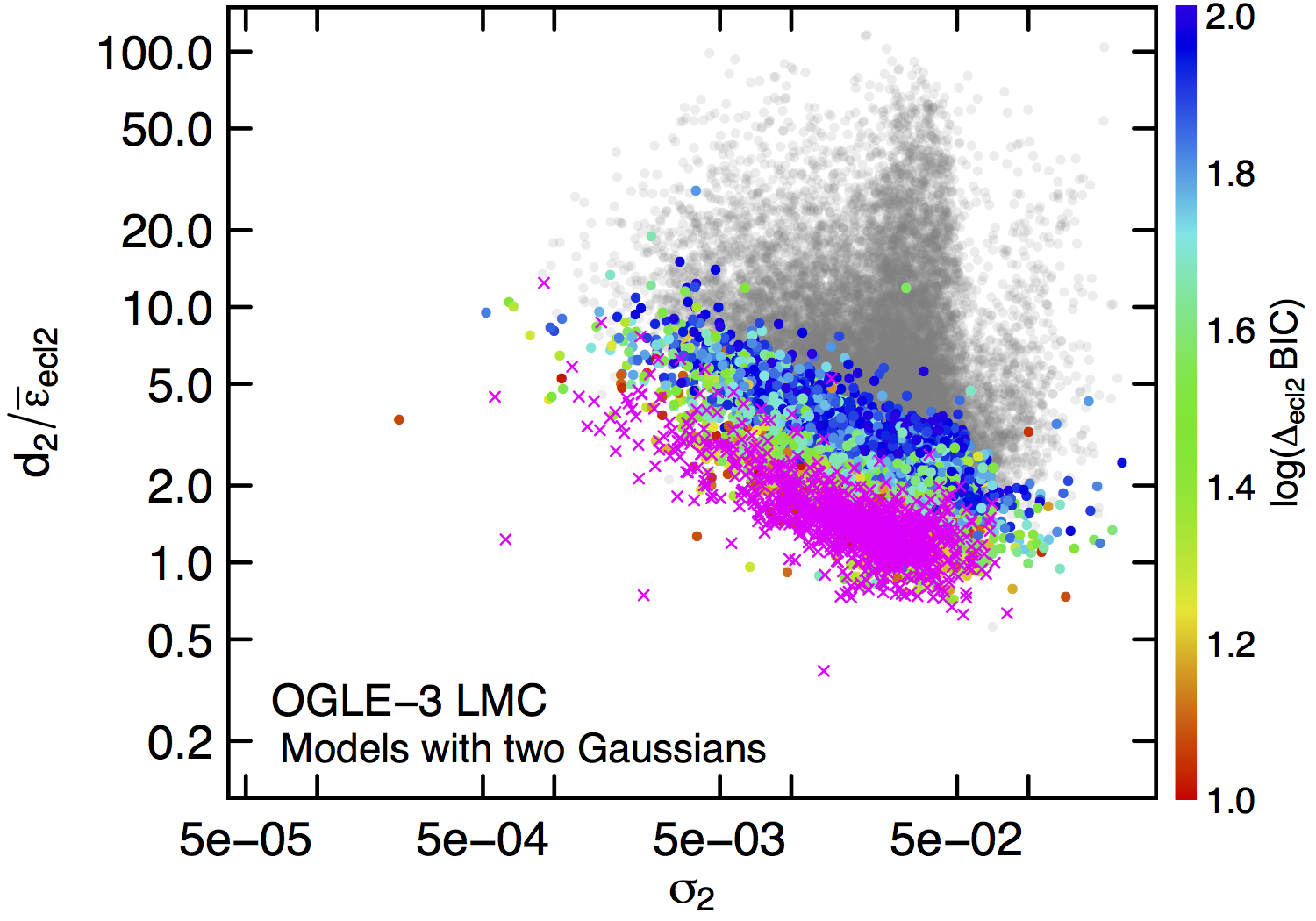}
  \caption{Same as Fig.~\ref{Fig:dOverMeanMagErrorVsSigma_ecl1}, but for secondary eclipse candidates of models containing two Gaussians.
           }
\label{Fig:dOverMeanMagErrorVsSigma_ecl2}
\end{figure}

\begin{figure}
  \centering
  \includegraphics[width=1.00\columnwidth]{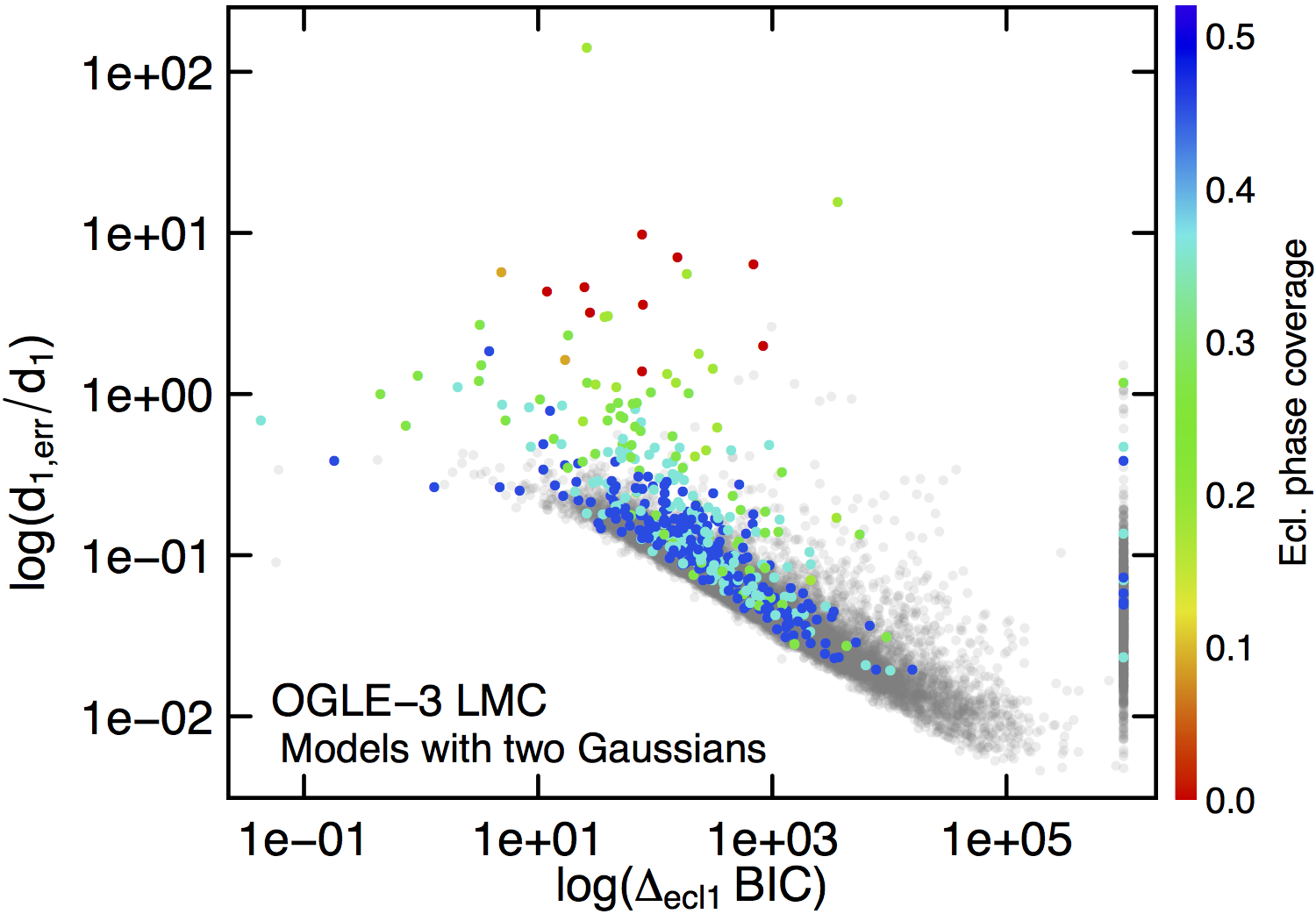}
  \caption{Relative uncertainty of the Gaussian depth of primary eclipse candidates versus eclipse significance for all models containing two Gaussians.
  The color indicates the eclipse coverage by the measurements according to the color scale shown on the right of the figure.
           A gray color is used for eclipse coverages larger than 50\%.
           }
\label{Fig:dRelErrVsEclSignificance_ecl1}
\end{figure}

It is instructive to further analyze the characteristics of the least reliable (according to $\Delta_\mathrm{ecl} \mathrm{BIC}$) eclipse candidates.
One expects those candidates to have Gaussian depth $d_i$ comparable to, or smaller than, the measurement uncertainties.
This is confirmed in Figs.~\ref{Fig:dOverMeanMagErrorVsSigma_ecl1} and \ref{Fig:dOverMeanMagErrorVsSigma_ecl2}, which show the ratio $d_i/\bar{\varepsilon}_{\mathrm{ecl}i}$ of the Gaussian depth over mean measurement uncertainty $\bar{\varepsilon}_{\mathrm{ecl}i}$ inside the eclipse versus Gaussian width $\sigma_i$ for primary ($i$=1) and secondary ($i$=2) eclipse candidates, respectively.
Sources that have $\Delta_{\mathrm{ecl}i} \mathrm{BIC}<100$ are shown in color in Figs.~\ref{Fig:dOverMeanMagErrorVsSigma_ecl1} and \ref{Fig:dOverMeanMagErrorVsSigma_ecl2}; they are seen to have the lowest $d_i/\bar{\varepsilon}_{\mathrm{ecl}i}$ ratios.

Figures~\ref{Fig:dOverMeanMagErrorVsSigma_ecl1} and \ref{Fig:dOverMeanMagErrorVsSigma_ecl2} further show a dependency of eclipse reliability on eclipse width.
Narrow eclipses require larger $d_i/\bar{\varepsilon}_{\mathrm{ecl}i}$ ratios than wide eclipses do in order to be significant, because narrow eclipses contain, on the mean, less measurements than wide eclipses.
Therefore, the narrower the eclipse is, the deeper it must be to be reliably detected.

A (small) fraction of eclipse candidates have Gaussian depth to mean measurement uncertainty ratios that are off the bulk distribution in the $d_i/\bar{\varepsilon}_{\mathrm{ecl}i}$ versus $\sigma_i$ diagrams shown in Figs.~\ref{Fig:dOverMeanMagErrorVsSigma_ecl1} and \ref{Fig:dOverMeanMagErrorVsSigma_ecl2}.
They concern very narrow eclipse candidates ($\sigma\lesssim 10^{-3}$) with $d_i/\bar{\varepsilon}_{\mathrm{ecl}i}$ ratios that can reach above 100.
They are mainly eclipse candidates that lack sufficient observations inside the eclipse.
As a result, the Gaussian depth cannot be well constrained, and an unrealistically deep Gaussian is adopted by the model fitting algorithm with a concomitant large uncertainty.
This is verified in Fig.~\ref{Fig:dRelErrVsEclSignificance_ecl1}, which shows the relative uncertainty $d_\mathrm{1,err}/d_1$ of the Gaussian depth versus eclipse significance $\Delta_\mathrm{ecl1} \mathrm{BIC}$ of the primary eclipse candidates of all models containing two Gaussians.
The eclipse coverage factor, shown in color for eclipses that have a coverage less than 50\%, is seen to be small for the models with small eclipse significance and/or with large relative uncertainty of the Gaussian depth.
They usually correspond to spurious eclipse candidates.

\subsubsection{Ellipsoidal component significance}
\label{Sect:modelsSignificance_ellAmplitude}

\begin{figure}
  \centering
  \includegraphics[width=1.00\columnwidth]{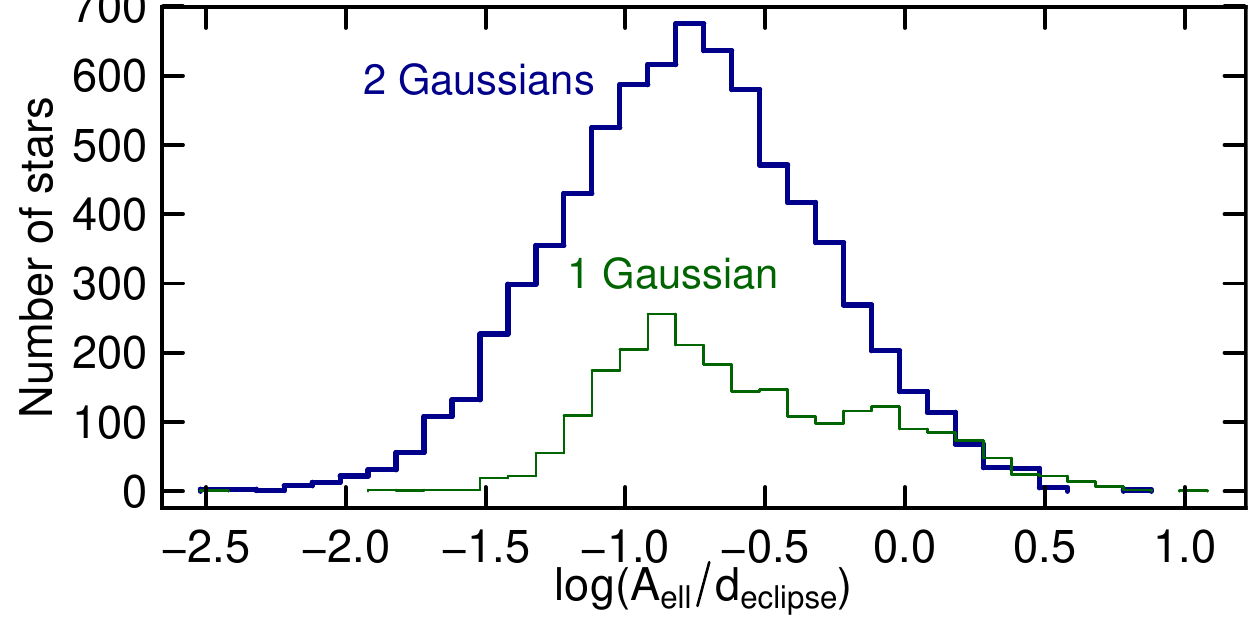}
  \caption{Distribution of the ratio of the ellipsoidal component amplitude over Gaussian depth $d_1$ of the primary eclipse candidate of models containing two Gaussians (thick blue histogram) or of models containing one Gaussian (thin green histogram).
          }
\label{Fig:histoAellOverD}
\end{figure}

\begin{figure}
  \centering
  \includegraphics[width=1.00\columnwidth]{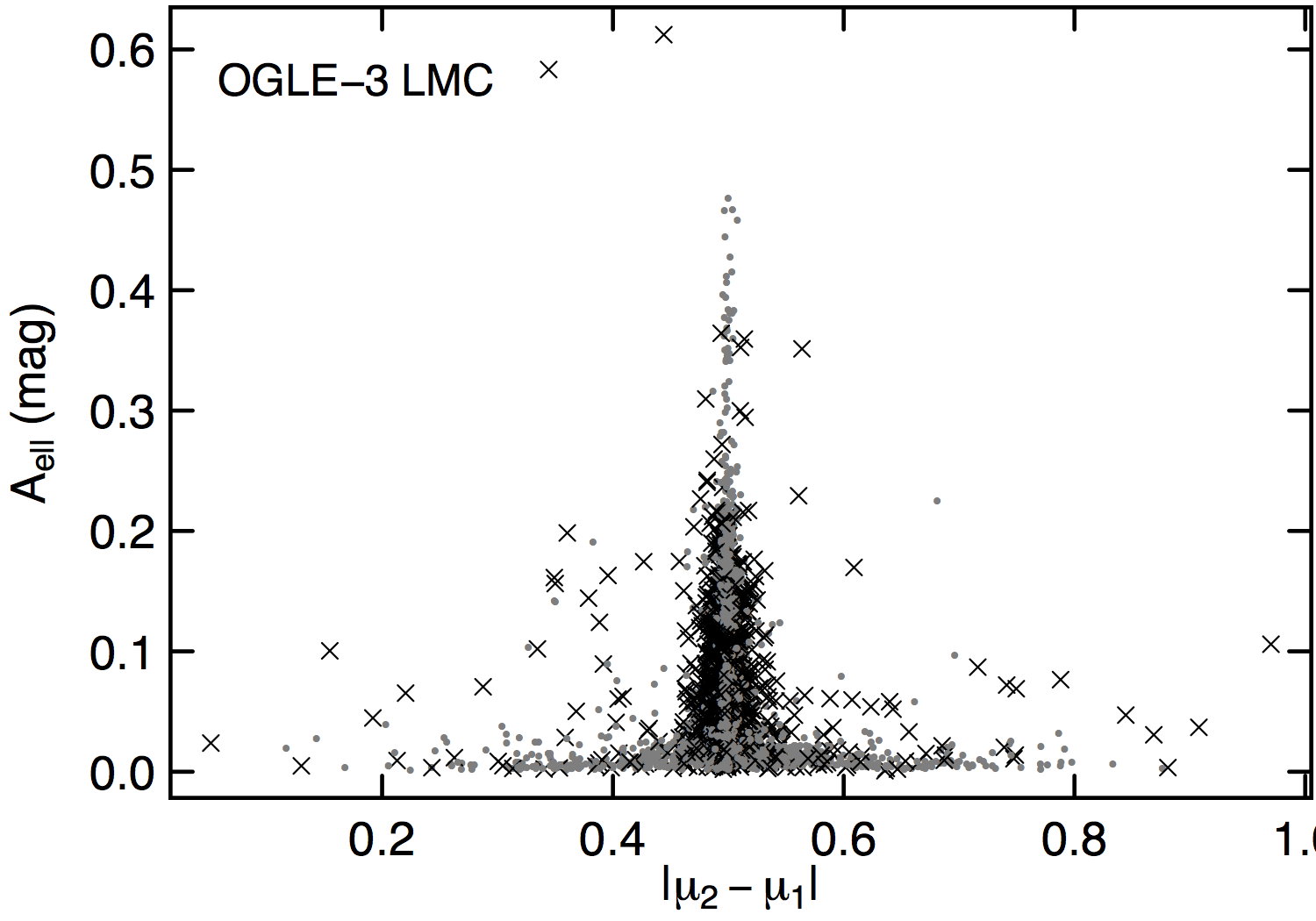}
  \caption{Amplitude of the ellipsoidal component versus the absolute phase difference between the locations of the two Gaussians.
  Models in which one of the two Gaussians has a BIC significance less than 50 are shown by cross markers.
           }
\label{Fig:AellDeltaMu}
\end{figure}

The amplitude $A_\mathrm{ell}$ of the ellipsoidal component can be relatively large compared to the Gaussian depth of the primary eclipse candidate.
The ratio $A_\mathrm{ell}/d_1$ is greater than 0.1 for \modif{69}\% (82\%) of models containing two (one) Gaussians, and greater than 0.5 for still \modif{15}\% (\modif{29}\%) of cases.
The histograms of the distributions of this ratio are shown in Fig.~\ref{Fig:histoAellOverD}.

The cosine function used in our models describes any type of ellipsoidal-like variability that would be present in the LCs.
For circular systems, the two eclipses are separated from each other by 0.5 in phase, and the ellipsoidal component, if present, is centered on both Gaussians (i.e. $\cos4\pi\mu_1=1$ and $\cos4\pi\mu_2=1$).
The case of elliptical systems containing two eclipses needs additional investigation.
For those systems, an ellipsoidal-like variability added to the two-Gaussian model would have the cosine centered on one of the two Gaussians and displaced relative to the other Gaussian.
Figure~\ref{Fig:AellDeltaMu} plots the amplitude of the ellipsoidal component versus phase separation $|\mu_2-\mu_1|$ between the two Gaussians of all CG12E1 and CG12E2 models.
It shows that the majority of models containing an ellipsoidal component have either a near-circular orbit ($|\mu_2-\mu_1|$ close to 0.5) or a small ellipsoidal component ($A_\mathrm{ell}<0.05$~mag).
Investigation of the few non-circular model candidates with a significant ellipsoidal component show that one of their Gaussians may be a spurious candidate ($\Delta_\mathrm{ecl1} \mathrm{BIC}$ or $\Delta_\mathrm{ecl2} \mathrm{BIC}<50$, shown with crosses in Fig.~\ref{Fig:AellDeltaMu}).

\subsection{Model degeneracies}
\label{Sect:modelsDegeneracies}

\begin{figure}
  \centering
  \includegraphics[width=1.00\columnwidth]{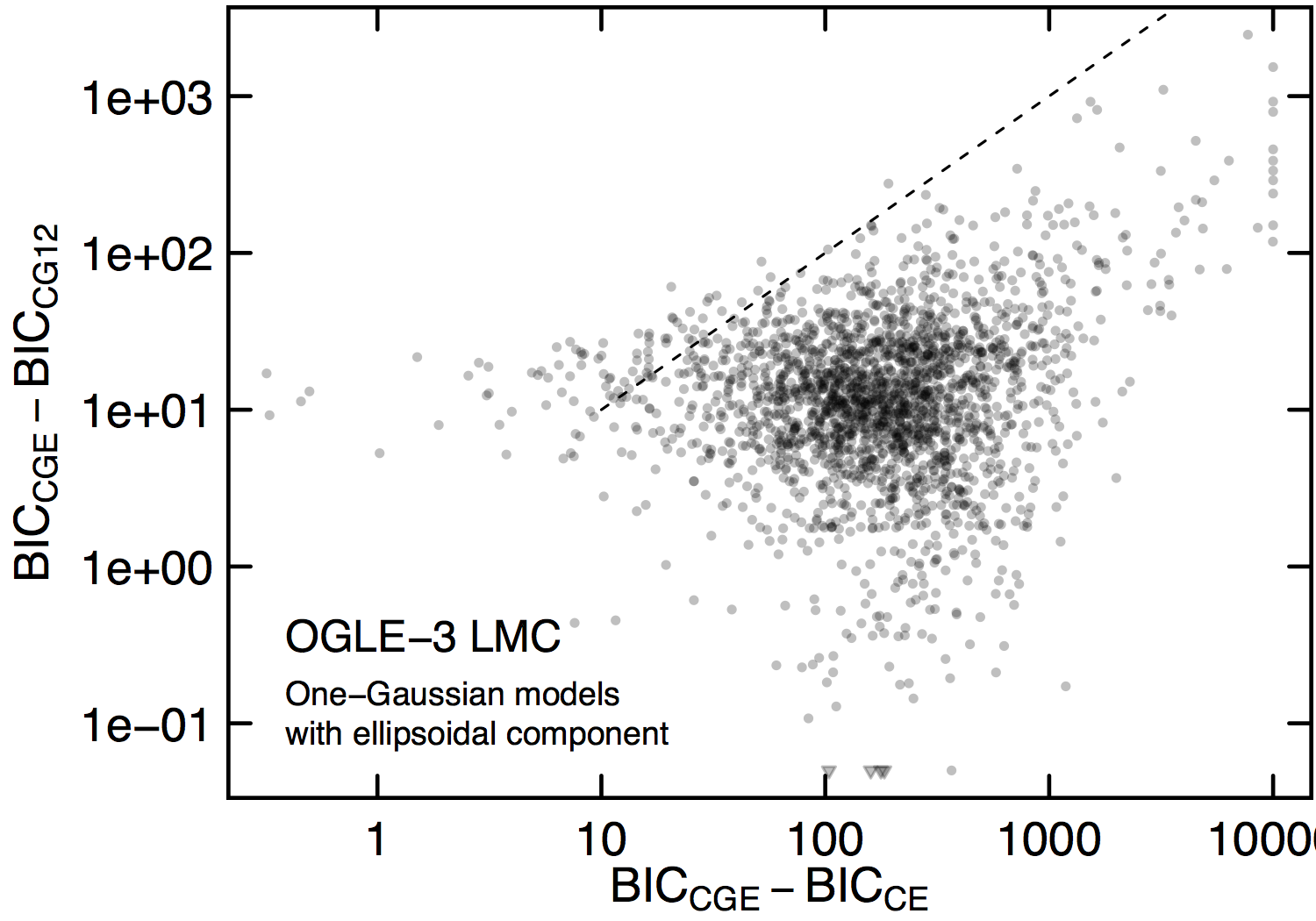}
  \caption{BIC value differences between CGE and CG12 models versus BIC value differences between CGE and CE models for all light curves for which the CGE model is favored from their BIC values.
           For clarity, the $\mathrm{BIC}_\mathrm{CGE} - \mathrm{BIC}_\mathrm{CG12}$ values in the figure are lower-bound limited to 0.05, and models having values lower than this limit are shown with downward triangles in the figure.
           A 45 degree diagonal dashed line is drawn as an eye-guide.
           }
\label{Fig:degeneracyCGE}
\end{figure}

The automated selection of a best model among several ones inevitably raises the question of model degeneracies given the data uncertainties.
In the case of two-Gaussian models, model degeneracy can arise because, for example, the cosine function describing the ellipsoidal-like variability can be mimicked by two wide Gaussian functions mirrored over adjacent phases.
Likewise, \textsl{EB}- and \textsl{EW}-type EBs can mathematically be modeled, within the measurement uncertainties, by either two wide Gaussians or an ellipsoidal component complemented by a wide Gaussian to account for the different eclipse depths.
Degenerate models are expected to have their BIC values close to one another, all of them describing almost equally well the data within the given measurement uncertainties.
We therefore estimate the degree of degeneracy between two models A and B by the absolute difference $|\mathrm{BIC}_\mathrm{A} - \mathrm{BIC}_\mathrm{B}|$ between their BIC values.

Figure~\ref{Fig:degeneracyCGE} illustrates model degeneracy for EBs that have the CGE model (one Gaussian + ellipsoidal component) selected by the automated procedure.
A CGE model can be degenerate with either a CG12 model (two Gaussians) or a CE model (only an ellipsoidal component).
The figure plots the degree of degeneracy of the CGE model with a CG12 model on the Y-axis versus the degree of degeneracy with a CE model on the X-axis.
Models located on the left part of the diagram (small $\mathrm{BIC}_\mathrm{CGE} - \mathrm{BIC}_\mathrm{CE}$) may be confused with a purely ellipsoidal model, while those in the lower part (small $\mathrm{BIC}_\mathrm{CGE} - \mathrm{BIC}_\mathrm{CG12}$) may be confused with a model containing only two Gaussians.
These degeneracies should be taken into consideration when performing statistical studies on an ensemble of EBs.
This exercise may be more or less straightforward depending on the type of degeneracy.
A degeneracy between CGE and CG models, for example, is not very harmful because CG models are equivalent to CGE models with zero amplitude of the ellipsoidal component.
A degeneracy between CGE and CG12 models, on the other hand, is more problematic.
In this case the alternative CG12 model would be composed of a narrow deep eclipse and a wide shallow eclipse, the astrophysical origin of which would be more challenging to find.

\subsection{Quality of the models}
\label{Sect:fitQuality}

\begin{figure}
  \centering
  \includegraphics[width=1.00\columnwidth]{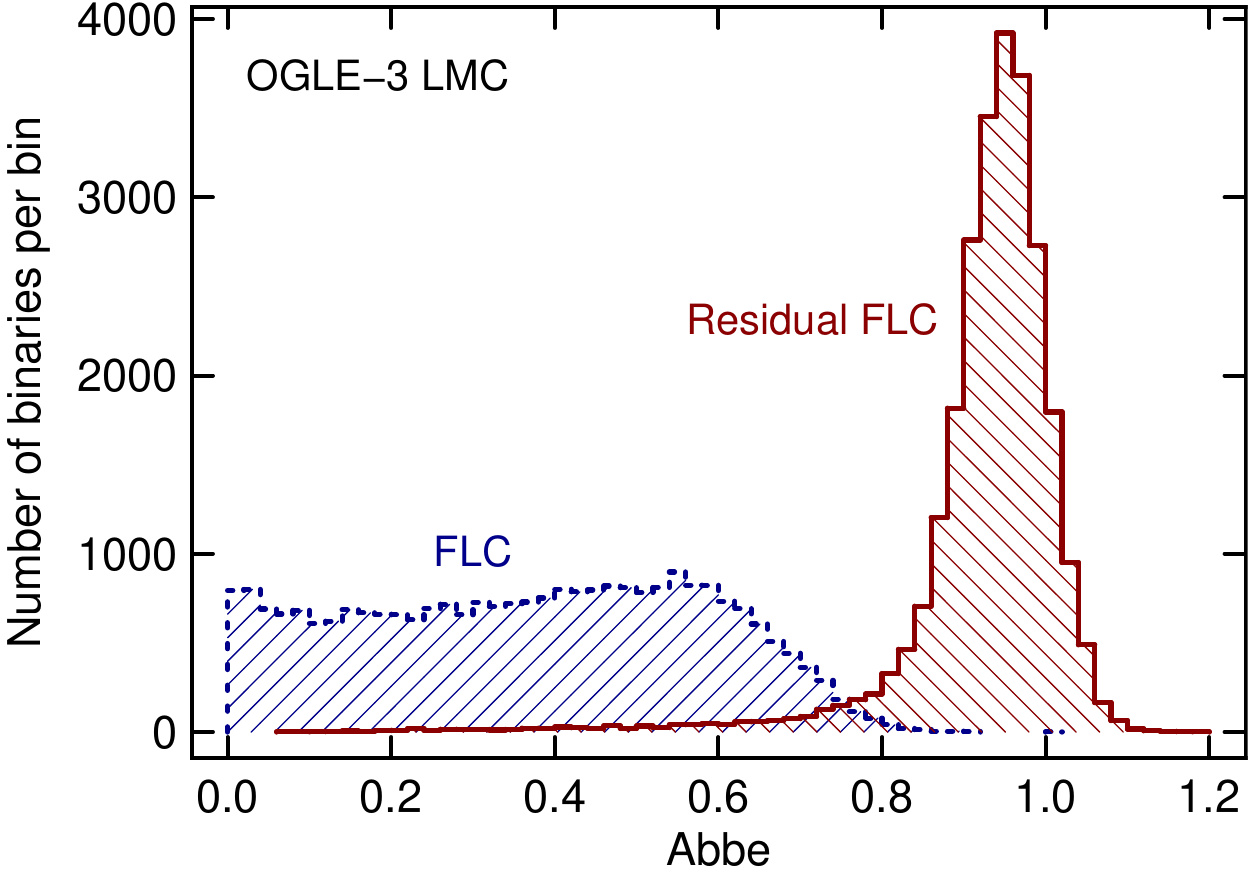}
  \caption{Histograms of the Abbe values of the original folded light curves (blue histogram with dashed contour and shaded at 45 degrees) and of the residual folded light curves (red histogram with solid contour and shaded at 135 degrees) of OGLE-III LMC eclipsing binaries.
           }
\label{Fig:histoAbbe}
\end{figure}

\begin{figure}
  \centering
  \includegraphics[width=1.00\columnwidth]{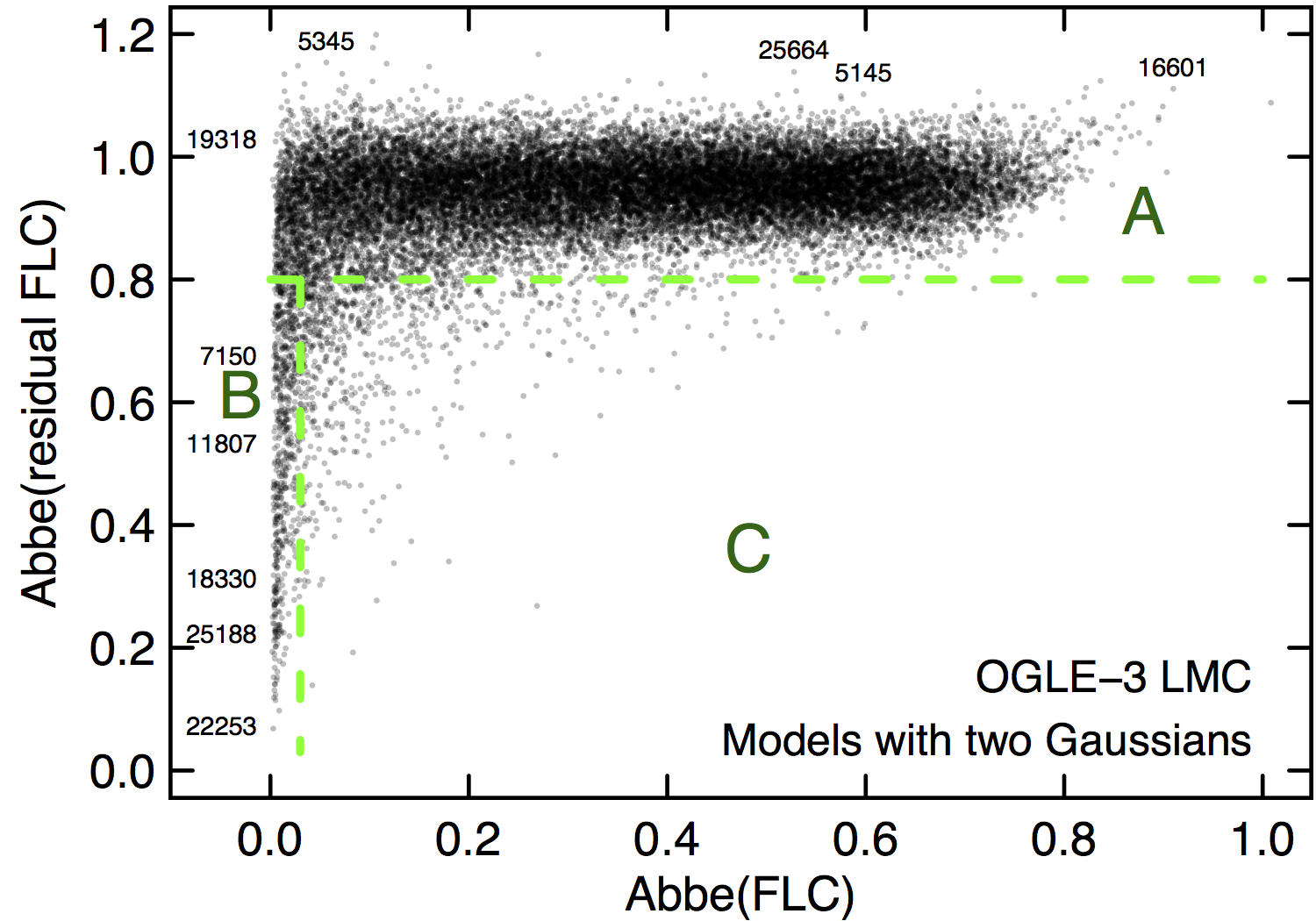}
  \caption{Abbe value $\Ab_\mathrm{resFLC}$ of the residual folded light curves versus Abbe value $\Ab_\mathrm{FLC}$ of the folded light curve of OGLE-III eclipsing binaries in the LMC, for all models containing two Gaussians (with or without ellipsoidal component).
           The dashed horizontal and vertical green lines delimit the three regions A, B and C mentioned in the text.
           Sources labelled in the figure have their folded light curves shown in Figs.~\ref{Fig:flcsRegionA_2G} to \ref{Fig:flcsRegionB_2G}.
           }
\label{Fig:AbbeVsAbbe_2G}
\end{figure}

We analyzed in the previous sections the significance of model components.
We now want to check how suitably the two-Gaussian models describe the variability patterns present in the FLCs of EBs.
To achieve this, we use the Abbe value that quantifies the degree of smooth variability present in a curve \citep[see][and references therein]{Mowlavi14}.
Given a series of $n$ values $y_{j=1\rightarrow n}$, the Abbe value $\Ab$ is defined by
\begin{equation}
  \Ab = \frac{n}{2(n-1)} \frac{\sum_{j=1}^{n-1}(y_{j+1}-y_j)^2}{\sum_{j=1}^{n}(y_j-\bar{y})^2} \,,
\label{Eq:Abbe}
\end{equation}
where $\bar{y}$ is the mean of $\{y_j\}$.

The Abbe values of the original and residual FLCs are noted $\Ab_\mathrm{FLC}$ and $\Ab_\mathrm{resFLC}$, respectively.
A FLC with no visible variability pattern will have an $\Ab_\mathrm{FLC}$ value around 1 (there is no correlation between successive $y_{j+1}-y_j$ differences), while a very clear and smooth variability pattern will result in a $\Ab_\mathrm{FLC}$ value decreasing to 0 ($y_{j+1}-y_j$ differences are small compared to the standard deviation of the series).
If a model successfully describes a FLC, no variability pattern should subsist in the residual FLC and the Abbe value $\Ab_\mathrm{resFLC}$ of the residual LC should be close to 1.

The histograms of $\Ab_\mathrm{FLC}$ and $\Ab_\mathrm{resFLC}$ are shown in Fig.~\ref{Fig:histoAbbe}.
The Abbe values of the original FLCs are seen to have an almost flat distribution between 0 and 0.6, and to start to decrease above 0.6.
This is expected, since an EB with $\Ab_\mathrm{FLC} \gtrsim 0.7$ is more difficult to be identified, and hence has a smaller probability to be in the OGLE-III catalog of EBs in the first place.
The Abbe values of the residual FLCs after model subtraction, on the other hand, peaks at 0.95 (thick red histogram in Fig.~\ref{Fig:histoAbbe}).
This reflects the efficiency of the two-Gaussian models to adequately fit the geometry of EB FLCs, thereby increasing the Abbe value from values below 0.8 in the original FLC to values above 0.8 in the residual FLC.
It does not guarantee, though, an adequate identification of eclipse and/or inter-eclipse components for the EB, which must be studied using component significances as done in the previous sections.
But it \modif{reveals} that no significant variability pattern remains in the FLC after model subtraction.

\begin{figure}
  \centering
  \includegraphics[width=1.00\columnwidth]{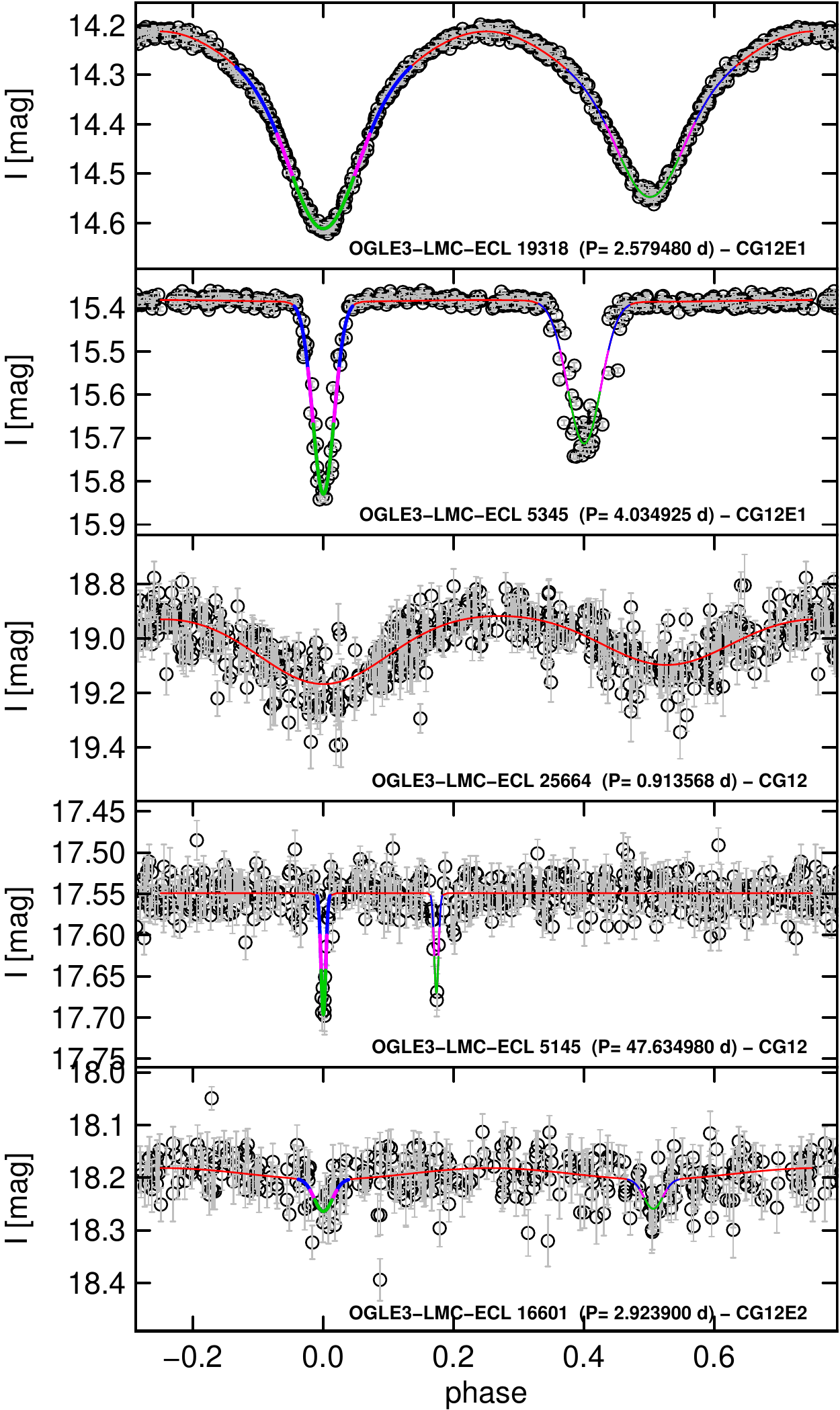}
  \caption{Examples of various folded light curves in region A ($\Ab_\mathrm{resFLC}>0.8$) of the $\Ab_\mathrm{resFLC}$ versus $\Ab_\mathrm{FLC}$ diagram.
           \modif{The colors of the models are the same as in Fig.~\ref{Fig:flcsSignificantComponents}.}
           Sources are ordered from top to bottom with increasing $\Ab_\mathrm{FLC}$ values, as labeled in Fig.~\ref{Fig:AbbeVsAbbe_2G}.
           }
\label{Fig:flcsRegionA_2G}
\end{figure}

\begin{figure}
  \centering
  \includegraphics[width=1.00\columnwidth]{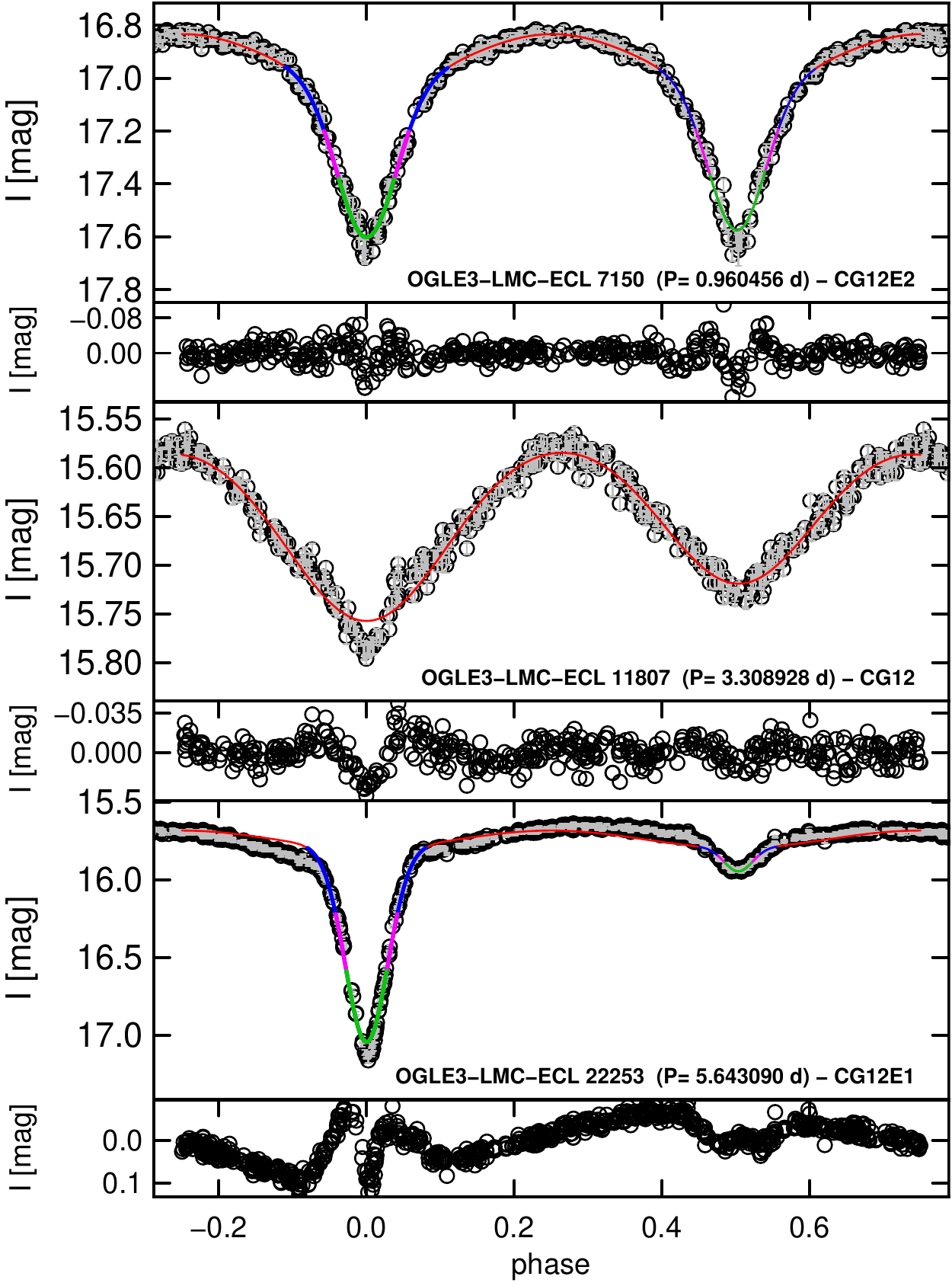}
  \includegraphics[width=1.00\columnwidth]{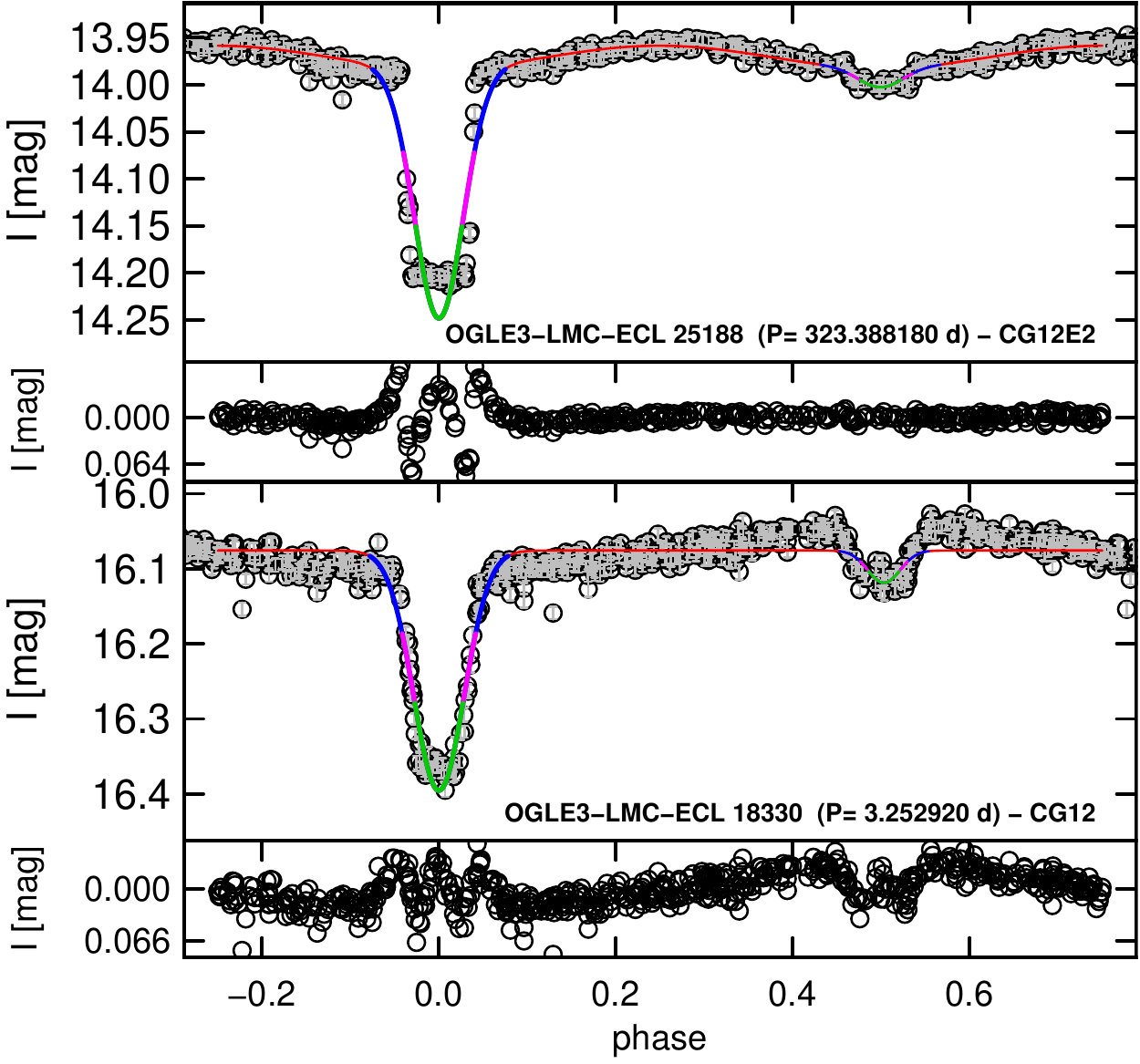}
  \caption{Examples of various folded light curves in region B of the $\Ab_\mathrm{resFLC}$ versus $\Ab_\mathrm{FLC}$ diagram, with the residual folded light curve of each source plotted in a smaller panel below the panel of each folded light curve.
           \modif{The colors of the models are the same as in Fig.~\ref{Fig:flcsSignificantComponents}.}
           The three top examples, ordered from top to down with increasing $\Ab_\mathrm{FLC}$ values, illustrate cases where Gaussian or cosine functions are not adequate enough to describe the light curve geometries of the eclipses or inter eclipses variability.
           The two bottom examples illustrate cases that require an additional physical effect not taken into account in the current two-Gaussian models.
           The upper example shows a case with a total eclipse, and the lower example a case with reflection.
           The positions of the sources in the $\Ab_\mathrm{resFLC}$ versus $\Ab_\mathrm{FLC}$ diagram are shown in Fig.~\ref{Fig:AbbeVsAbbe_2G}.
           }
\label{Fig:flcsRegionB_2G}
\end{figure}

A small fraction of EBs have a residual Abbe value below 0.8, as shown by the tail distribution of $\Ab_\mathrm{resFLC}$ in Fig.~\ref{Fig:histoAbbe}.
In those cases, a variability pattern that can be significant remains in the FLC after model subtraction.
To further analyze these cases, the Abbe value $\Ab_\mathrm{resFLC}$ is plotted versus $\Ab_\mathrm{FLC}$ in Fig.~\ref{Fig:AbbeVsAbbe_2G} for all models containing two Gaussians (irrespective of the presence of an ellipsoidal component).
Three regions are identified in the diagram:
\begin{itemize}
\item Region A ($\Ab_\mathrm{resFLC} > 0.8$):
the two-Gaussian model reproduces well the geometry of the FLCs within the measurement uncertainties.
This represents by far the majority of cases, with 94\% of all OGLE-III EB LCs falling in this region.
The FLCs of several examples labeled in Fig.~\ref{Fig:AbbeVsAbbe_2G} are shown in Fig.~\ref{Fig:flcsRegionA_2G} with, from top to bottom panel, increasing $\Ab_\mathrm{FLC}$ (i.e. decreasing LC signal-to-noise ratio).
\vskip 1mm
\item Region B ($\Ab_\mathrm{resFLC} < 0.8$, $\Ab_\mathrm{FLC} < 0.03$): the signal-to-noise of the LCs is very high (with a resulting $\Ab_\mathrm{FLC} < 0.03$), and the well defined FLC geometry challenges the two-Gaussian model (as seen from the variability pattern still present in the residual FLC, with $\Ab_\mathrm{resFLC} < 0.8$).
Only 2.4\% of sources modeled with two Gaussians fall in this region of the $\Ab_\mathrm{resFLC}$ versus $\Ab_\mathrm{FLC}$ diagram.
The two-Gaussian model can fail to adequately describe the FLC geometry for two reasons.
First, the LC geometries during the eclipse and inter eclipse phases are more complex than what can be described by simple Gaussian and cosine functions, respectively. 
The three top FLCs in Fig.~\ref{Fig:flcsRegionB_2G} illustrate such cases, with the residual FLCs displayed in a panel below each FLC.
Nevertheless, the examples show that the two-Gaussian models still successful grasp the main properties of the eclipses despite the simplicity of the models.

The two-Gaussian model can also fail to adequately describe the geometry of an EB FLC if a physical effect other than an eclipse or ellipsoidal-like variability is present in the LC.
This is the case, for example, if the system has a total eclipse or a reflection component.
An example of each of those two cases occurring in region B is shown in the bottom panels of Fig.~\ref{Fig:flcsRegionB_2G}.
\vskip 1mm
\item Region C ($\Ab_\mathrm{resFLC} < 0.8$, $\Ab_\mathrm{FLC} > 0.03$): sources in this region should be successfully modeled by a two-Gaussian model\modif{, because the relatively low S/N is less demanding of the model than in region B}.
Failure of the two-Gaussian model to do so would imply either additional physics not accounted for in the two-Gaussian model, or more fundamental issues to be investigated.
This region contains thus potentially interesting cases of outliers to be investigated.
They will be addressed in Sect.~\ref{Sect:discussion_outliers_Abbe}.
\end{itemize}

\begin{figure}
  \centering
  \includegraphics[width=1.00\columnwidth]{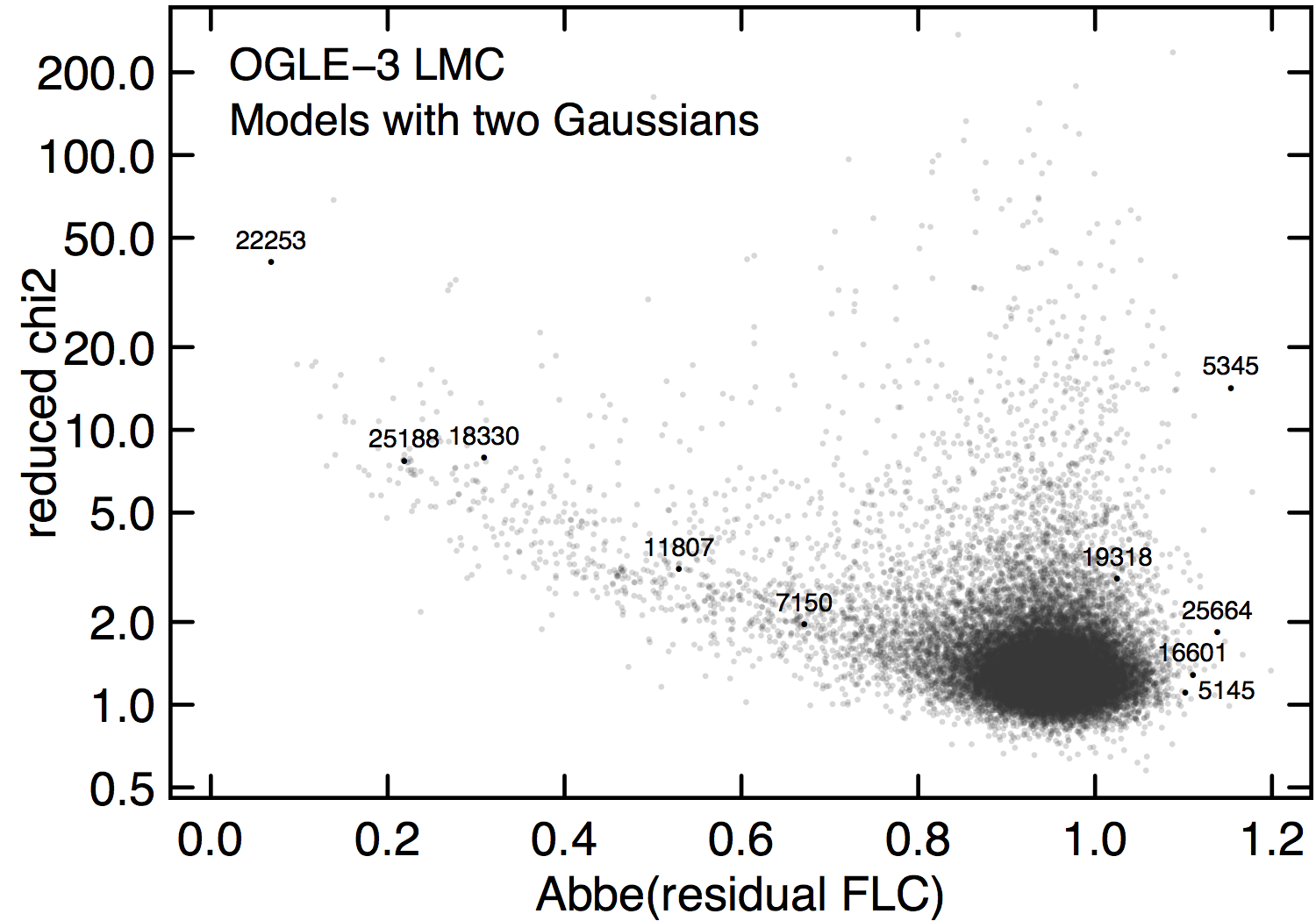}
  \caption{Reduced $\chi^2$ versus Abbe value $\Ab_\mathrm{resFLC}$ of the residual folded light curve for all models containing two Gaussians (with or without ellipsoidal component).
           Labelled sources are the ones that are labelled in Fig.~\ref{Fig:AbbeVsAbbe_2G}.
           }
\label{Fig:reducedChi2VsAbbe_2G}
\end{figure}

Sources falling in region C of the $\Ab_\mathrm{resFLC}$ versus $\Ab_\mathrm{FLC}$ diagram are expected to have a large dispersion in their residual LCs.
Therefore, their reduced $\chi^2$, defined by
\begin{equation}
  \chi_\mathrm{reduced}^2 = \frac{1}{(N_\mathrm{obs}-p)} \sum_{i=1}^{N_\mathrm{obs}} \frac{\left[y_i(\varphi_i) - G_i(\varphi_i)\right]^2}{\varepsilon_i^2}
\end{equation}
where $p$ is the number of parameters in the model and $\varepsilon_i$ is the uncertainty on the magnitude of measurement $i$, should be large.
Figure~\ref{Fig:reducedChi2VsAbbe_2G} plots $\chi_\mathrm{reduced}^2$ versus $\Ab_\mathrm{resFLC}$.
The majority of sources are seen in the figure to have $\Ab_\mathrm{resFLC} \simeq 0.95$ and $\chi_\mathrm{reduced}^2 \simeq 1.5$.
A stream of sources is also seen towards lower $\Ab_\mathrm{resFLC}$ values with a correlated increase of $\chi_\mathrm{reduced}^2$.
This is expected, since values of $\Ab_\mathrm{resFLC}$ smaller than 0.7 indicate the presence of residual variability patterns that result in larger $\chi_\mathrm{reduced}^2$ values.
And indeed, the sources labelled in region C of Fig.~\ref{Fig:AbbeVsAbbe_2G} lie on or close to this stream of data points in Fig.~\ref{Fig:reducedChi2VsAbbe_2G}.

Figure~\ref{Fig:reducedChi2VsAbbe_2G}, however, shows the existence of a subset of sources that have $\chi_\mathrm{reduced}^2$ larger than what is expected from the bulk or stream distributions of points in the figure.
They indicate the presence of an additional variability component of a different nature, that breaks the smoothly varying pattern of a FLC derived from a strictly periodic variability.
These cases will further be studied in Sect.~\ref{Sect:discussion_outliers_chi2}.

\subsection{Table summary}
\label{Sect:tableSummary}

All quantities derived in this study for the OGLE-III LMC EBs are published in a table available in electronic format.
A description of the table content is given in Appendix~\ref{Appendix:tableDescription}.

\section{Discussion}
\label{Sect:discussion}

We present two application examples of the two-Gaussian models.
The first one (Sect.~\ref{Sect:discussion_outliers}) aims to identify binary systems in physical configurations incompatible with two-Gaussian models.
We refer to these systems as outliers.
The second example (Sect.~\ref{Sect:discussion_statisticalAnalysis}) shows how the two-Gaussian model results can be used to study statistical properties of the ensemble of EBs.
They are given here for illustrative purposes only, a full study of each of these two applications is beyond the scope of this paper.

\subsection{Identification of outlying cases}
\label{Sect:discussion_outliers}

The choice of Gaussian and cosine functions to model the FLC geometry of eclipse and ellipsoidal variability, respectively, defines the set of EB configurations than can be described by the two-Gaussian models.
Any deviation from this set of configurations will be detectable through poor model fit quality.
We use here the two diagnostic tools presented in Sect.~\ref{Sect:fitQuality} to evaluate model fit quality: the $\Ab_\mathrm{resFLC}$ versus $\Ab_\mathrm{FLC}$ diagram (Fig.~\ref{Fig:AbbeVsAbbe_2G}) and $\chi_\mathrm{reduced}^2$ versus $\Ab_\mathrm{resFLC}$ diagram (Fig.~\ref{Fig:reducedChi2VsAbbe_2G}).
We discuss these two diagrams in Sects.~\ref{Sect:discussion_outliers_Abbe} and \ref{Sect:discussion_outliers_chi2}, respectively.

\subsubsection{Outliers from the $\Ab_\mathrm{resFLC}$ versus $\Ab_\mathrm{FLC}$ diagram}
\label{Sect:discussion_outliers_Abbe}

\begin{figure*}
  \centering
  \includegraphics[width=2.00\columnwidth]{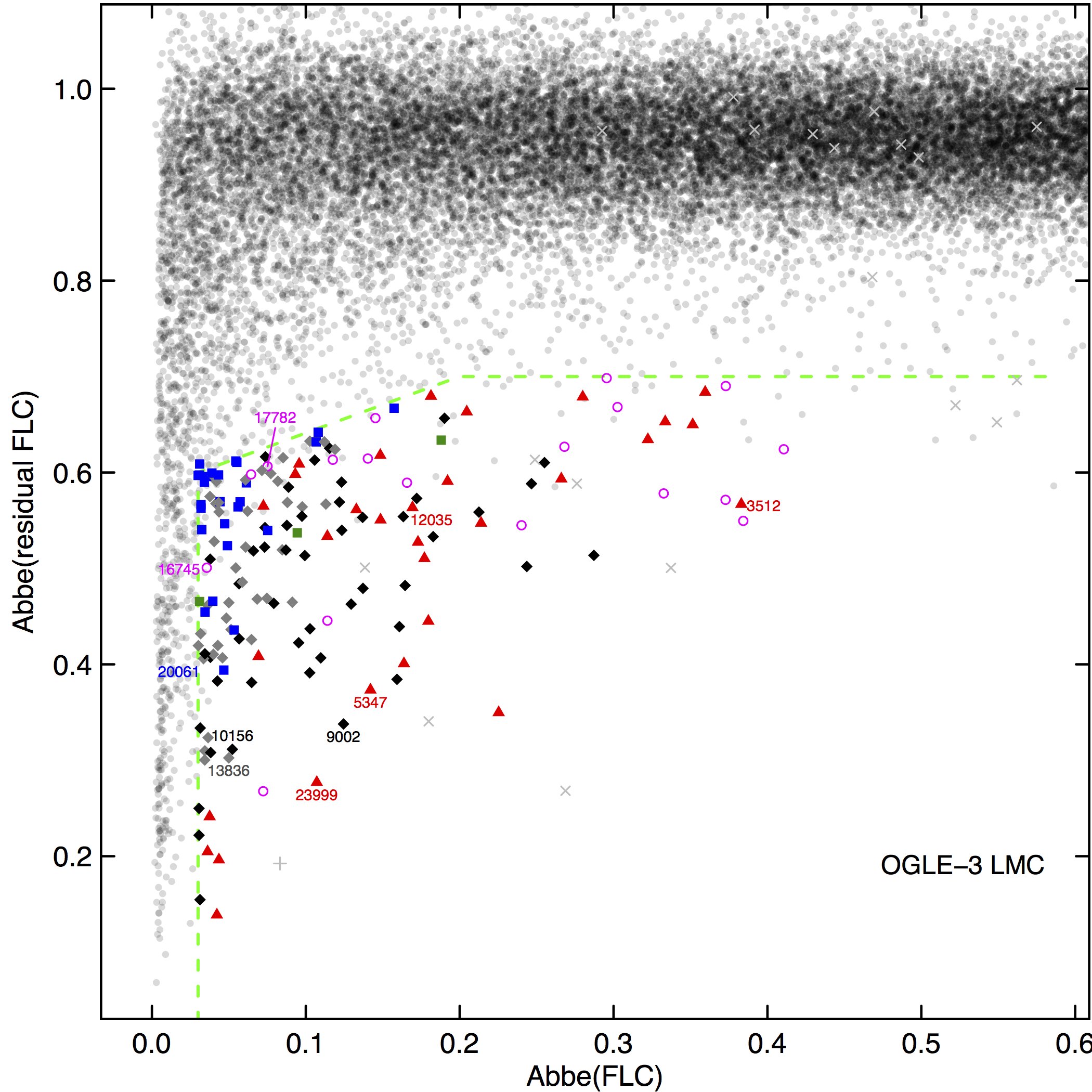}
  \caption{Same as Fig.~\ref{Fig:AbbeVsAbbe_2G}, but for all two-Gaussian models irrespective of the number of Gaussians, and zoomed in Region C of the $\Ab_\mathrm{resFLC}$ versus $\Ab_\mathrm{FLC}$ diagram.
           Sources in the lower-right area delimited by the green dashed line have been visually classified into one of the following types of eclipsing binary:
           systems showing a total eclipse (blue filled squares),
           systems with semi-detached morphology (gray filled diamonds),
           systems with a reflection-like effect (black filled diamonds),
           systems with eccentric tidal distortions (red filled triangles),
           systems with other special effects (magenta open circles),
           systems for which the two-Gaussian model procedure failed to identify at least one eclipse in the folded light curve (gray crosses) or of which the orbital period is wrong (gray plus sign).
           Sources labelled in the figure have their folded light curves shown in Figs.~\ref{Fig:flcsRegionC} and \ref{Fig:flcsRegionCIntrinsicScatter}.
           }
\label{Fig:AbbeVsAbbe_RegionC}
\end{figure*}

\begin{figure}
  \centering
  \includegraphics[width=0.975\columnwidth]{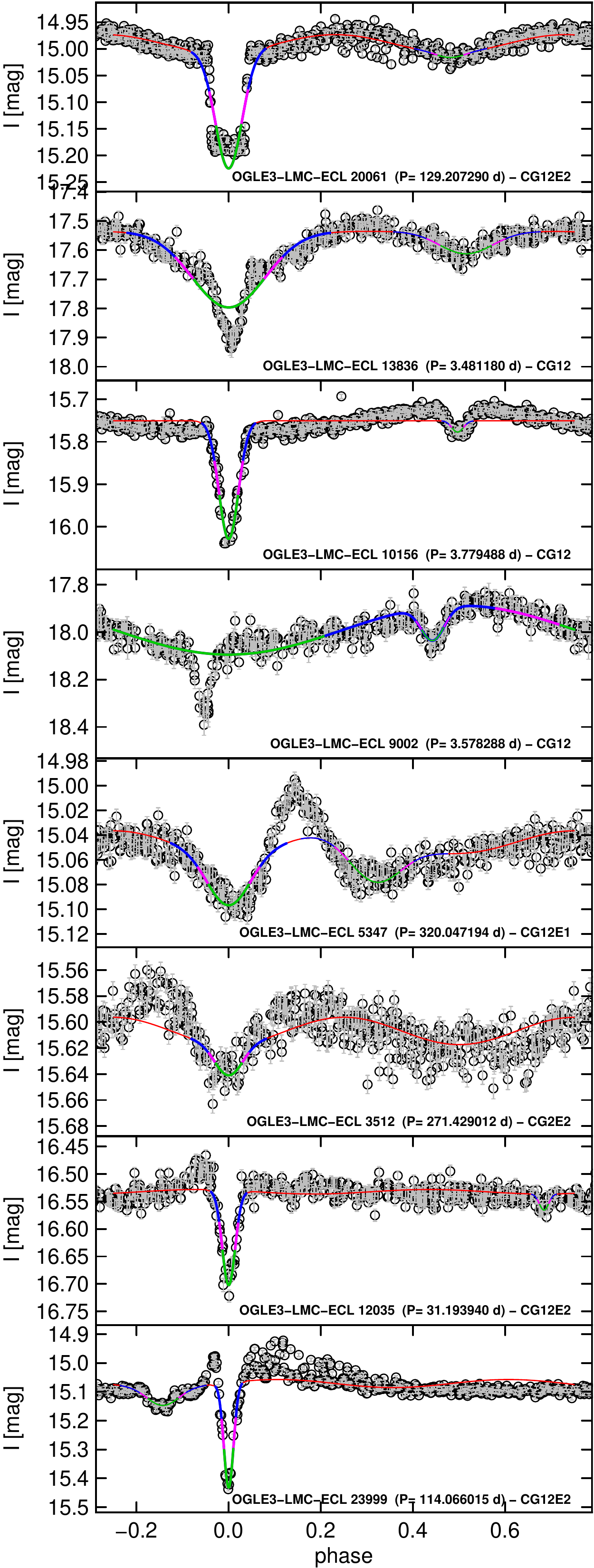}
  \caption{Examples of various folded light curvess in region C of the $\Ab_\mathrm{resFLC}$ versus $\Ab_\mathrm{FLC}$ diagram.
           \modif{The colors of the models are the same as in Fig.~\ref{Fig:flcsSignificantComponents}.}
           From top to bottom: a case with a total eclipse, a case with a semi-detached morphology, two cases with reflection-like effect, and four cases with eccentric tidal distortions.
  Their positions in the $\Ab_\mathrm{resFLC}$ versus $\Ab_\mathrm{FLC}$ diagram are labelled in Fig.~\ref{Fig:AbbeVsAbbe_RegionC}.
           }
\label{Fig:flcsRegionC}
\end{figure}

\begin{figure}
  \centering
  \includegraphics[width=1.00\columnwidth]{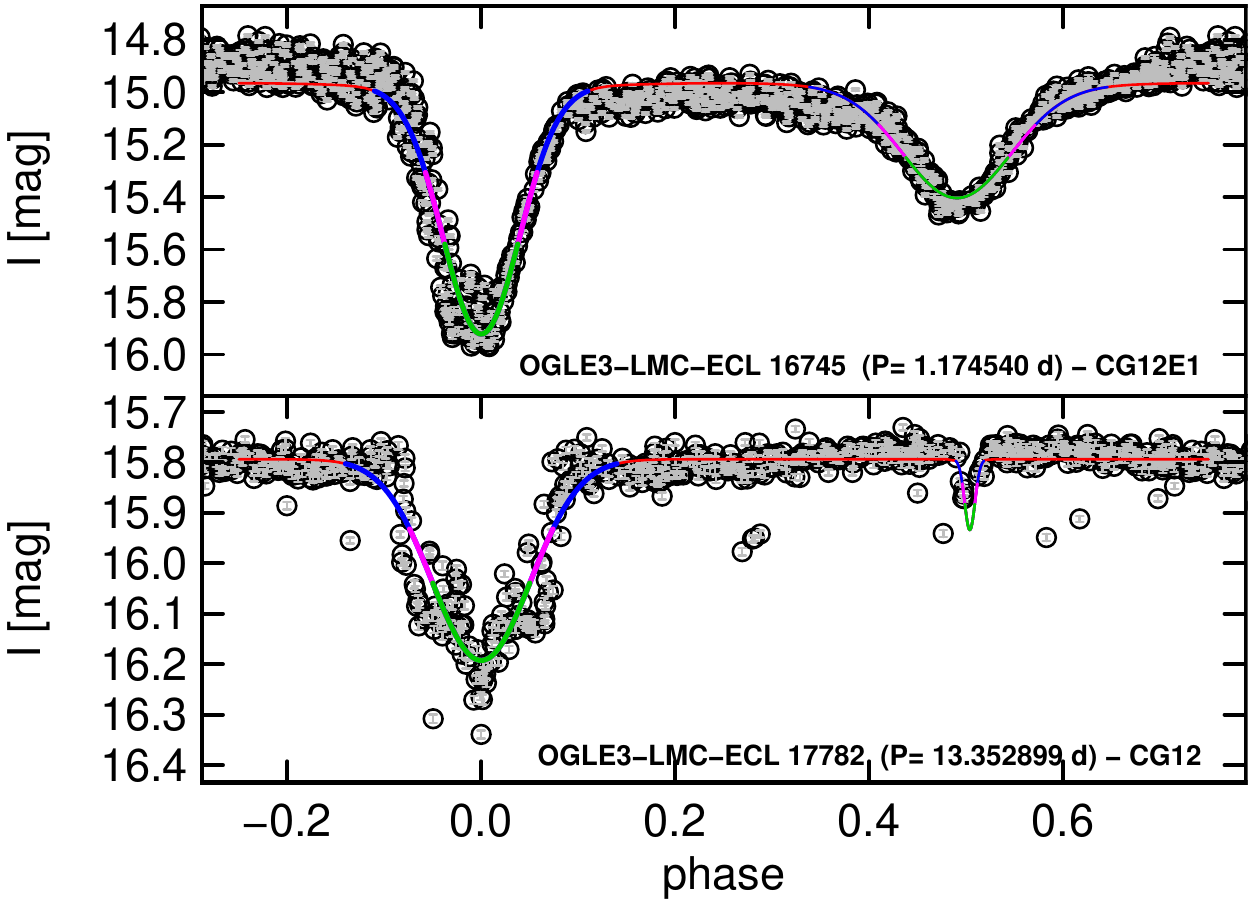}
  \caption{Examples of folded light curves in region C of the $\Ab_\mathrm{resFLC}$ versus $\Ab_\mathrm{FLC}$ that have large intrinsic scatter in their residuals.
           \modif{The colors of the models are the same as in Fig.~\ref{Fig:flcsSignificantComponents}.}
           Their positions in the $\Ab_\mathrm{resFLC}$ versus $\Ab_\mathrm{FLC}$ diagram are labelled in Fig.~\ref{Fig:AbbeVsAbbe_RegionC}.
           }
\label{Fig:flcsRegionCIntrinsicScatter}
\end{figure}

We select all systems from region C of the $\Ab_\mathrm{resFLC}$ versus $\Ab_\mathrm{FLC}$ diagram (Fig.~\ref{Fig:AbbeVsAbbe_2G}).
A zoomed version of the figure is shown in Fig.~\ref{Fig:AbbeVsAbbe_RegionC}.
The FLCs of all sources that lie within the lower-right area delimited by the dashed green line in Fig.~\ref{Fig:AbbeVsAbbe_RegionC} have been visually inspected and classified in one of the categories described below.
Examples are provided in Figs.~\ref{Fig:flcsRegionC} and \ref{Fig:flcsRegionCIntrinsicScatter}.

\paragraph{Total eclipse}
(blue filled squares in Fig.~\ref{Fig:AbbeVsAbbe_RegionC}) --
The presence of a total eclipse manifests itself by a flat bottom in the LC during the eclipse.
This is poorly approximated by a Gaussian function, and will result in an $\Ab_\mathrm{resFLC}$ value below 0.8.
An example is given with source 20061 in Fig.~\ref{Fig:flcsRegionC}.
Systems observed with close-to-total eclipses will also have LC geometry that deviates from Gaussian, because of the steep ingress and egress curves.
A limiting case is given by a system containing two stars of equal radii in circular orbit, for which the LC during the eclipse will have a V-shaped geometry if the system is seen edge-on.

\paragraph{Semi-detached morphology}
(gray filled diamonds in Fig.~\ref{Fig:AbbeVsAbbe_RegionC}) --
Systems that have one of the stars filling or close to filling its Roche lobe will display a non-cosine LC shape between the eclipses.
An example is given with source 13836 shown in Fig.~\ref{Fig:flcsRegionC}.

\paragraph{Reflection effect}
(black filled diamonds in Fig.~\ref{Fig:AbbeVsAbbe_RegionC}) --
The LCs of some systems show an out-of-eclipse brightening around the secondary eclipse.
An example is given with sources 10156 and 9002 in Fig.~\ref{Fig:flcsRegionC}.
This can be due to reflection, where the hotter star heats the surface of the cooler star that faces the hot star.
For source 9002, a lag is visible between the phase of the secondary eclipse and the phase at maximum luminosity, which could be caused by stellar rotation.
We classify these systems as having a reflection signature in their LC.
Their LCs cannot be modeled with a cosine function with half the orbital period used to model ellipsoidal-like variability, but could successfully be described with a cosine function with a period equal to the orbital period \citep{MoeDiStefano15}.

\paragraph{Tidal distortions in eccentric binaries and heartbeat stars}
(red filled triangles in Fig.~\ref{Fig:AbbeVsAbbe_RegionC}) --
The effect on the LC of tidal distortions in eccentric binaries also appears in region C of the $\Ab_\mathrm{resFLC}$ versus $\Ab_\mathrm{FLC}$ diagram.
The FLCs of four cases showing LC deformations due to such effects are given in Fig.~\ref{Fig:flcsRegionC} with sources 5347, \modif{3512}, 12035 and 23999.
Various LC geometries due to tidal distortions have been reported by \cite{ThompsonEverettMullally_etal12} in the \textsl{Kepler} data.

\paragraph{Large intrinsic scatter}
(magenta open circles in Fig.~\ref{Fig:AbbeVsAbbe_RegionC}) --
Some systems display a scatter in their residual LC larger than what is expected from the measurement uncertainties.
Two such cases are shown in Fig.~\ref{Fig:flcsRegionCIntrinsicScatter} with sources 16745 and 17782.
They are further discussed in Sect.~\ref{Sect:discussion_outliers_chi2}.

\paragraph{Failed convergence}
(gray cross and plus signs in Fig.~\ref{Fig:AbbeVsAbbe_RegionC}) --
The mismatch results from either a failure to correctly identify the initial locations of the eclipses or to converge on the 2-Gaussian model (gray crosses in Fig.~\ref{Fig:AbbeVsAbbe_RegionC}), or from a wrong initial orbital period (gray plus sign in Fig.~\ref{Fig:AbbeVsAbbe_RegionC}).
Only one clear case of the last category is found in the OGLE-III catalogue of LMC EBs, for which the double of the true period is reported in the OGLE-III catalogue.

\subsubsection{Outliers in the $\chi_\mathrm{reduced}^2$ versus $\Ab_\mathrm{resFLC}$ diagram}
\label{Sect:discussion_outliers_chi2}

\begin{figure*}
  \centering
  \includegraphics[width=2.00\columnwidth]{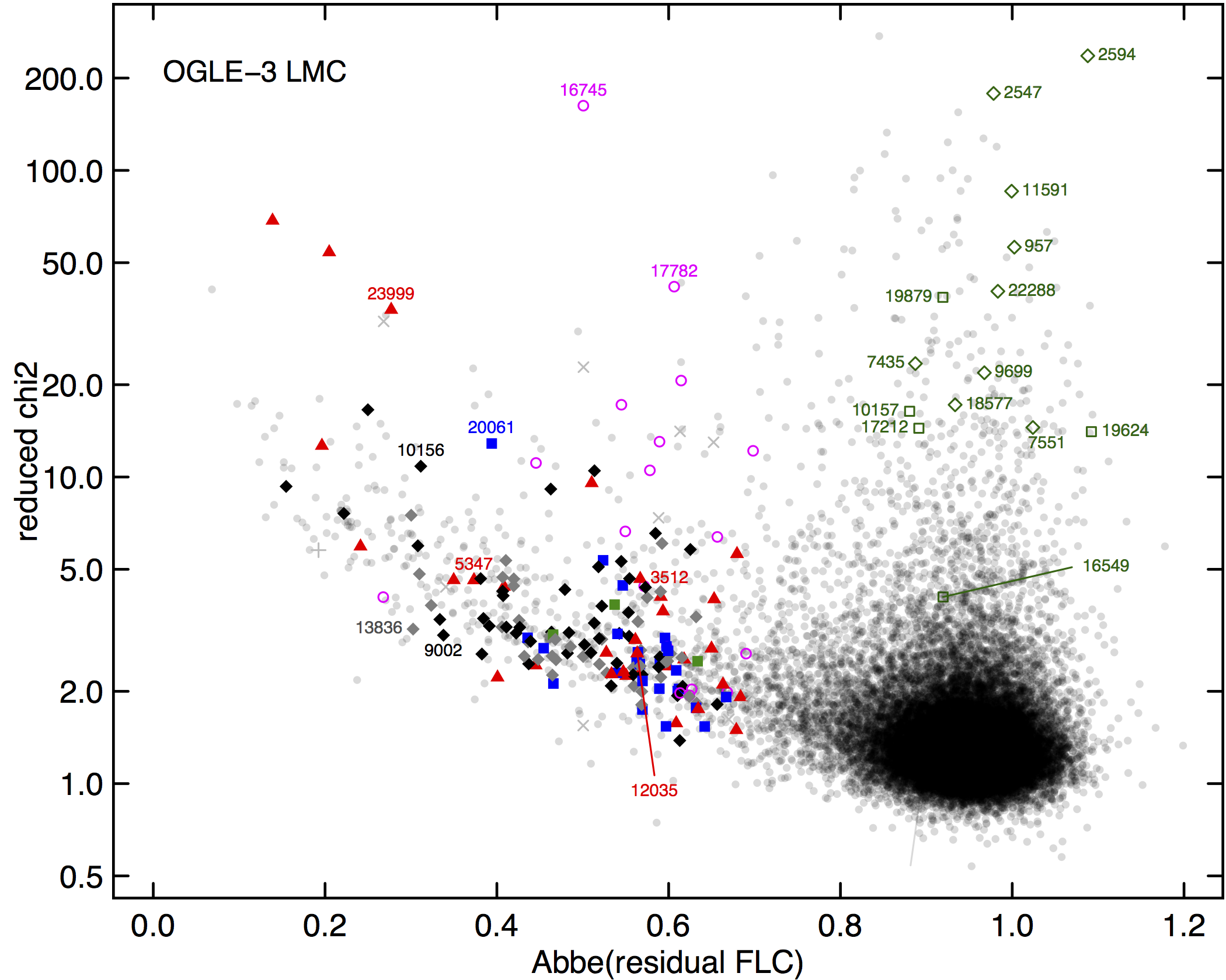}
  \caption{Same as Fig.~\ref{Fig:reducedChi2VsAbbe_2G}, but for all two-Gaussian models irrespective of the number of Gaussians.
           Labeled sources with $\Ab_\mathrm{resFLC}<0.8$ have their folded light curves shown in Figs.~\ref{Fig:flcsRegionC}, \ref{Fig:flcsRegionCIntrinsicScatter}.
           Labeled sources plotted with an open diamond have their light curves and folded light curves shown in Fig.~\ref{Fig:flcsInstrinsicScatter}.
           Labeled sources plotted with an open square have their folded light curves shown in Figs.~\ref{Fig:flcsPerturbedEclipse} and \ref{Fig:flcsMultipleSystem}.
           }
\label{Fig:reducedChi2VsAbbe_outliers}
\end{figure*}

\begin{figure*}
  \centering
  \includegraphics[width=0.88\columnwidth]{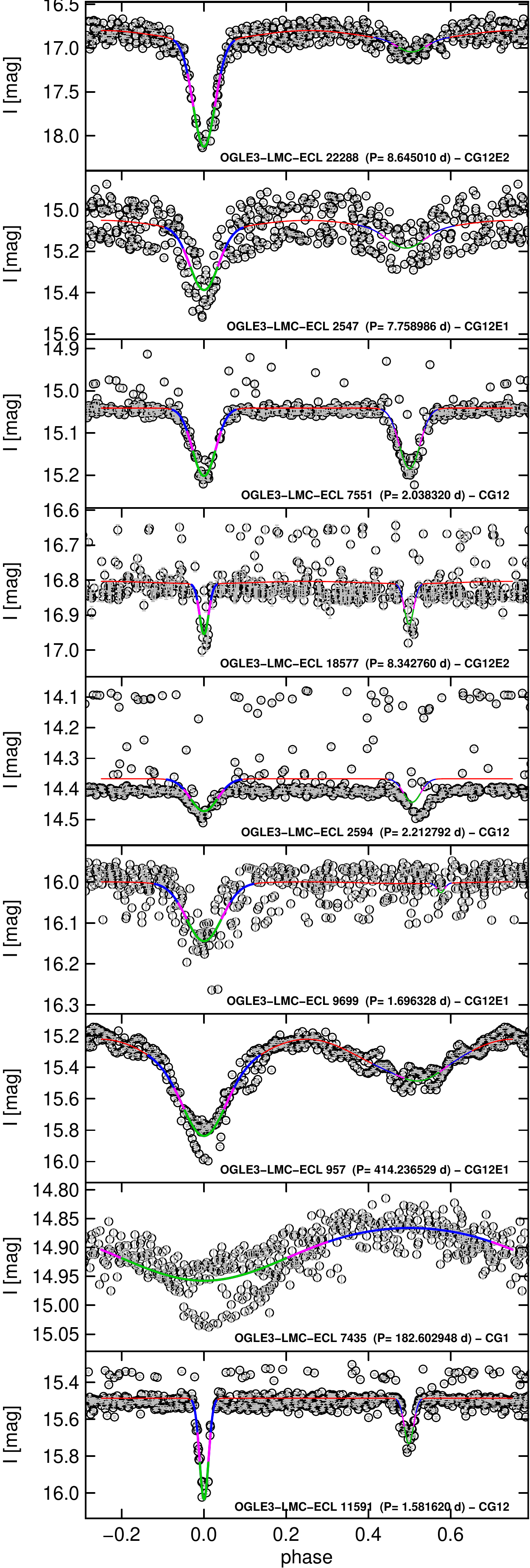}
  \includegraphics[width=0.88\columnwidth]{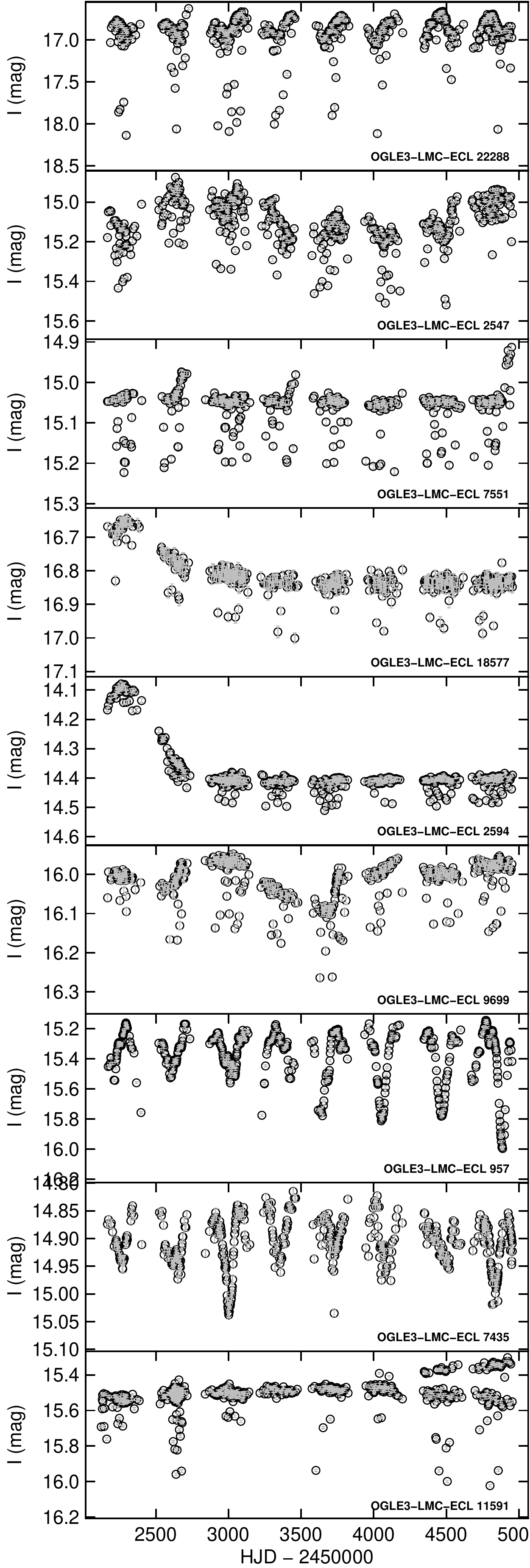}
  \caption{Examples of various folded light curves (left plots) and their light curves (right plots) having a large reduced $\chi^2$.
           From top to bottom: two cases with intrinsic quasi-periodic variability, one case showing flares, two cases with an outburst, one irregular variable, two cases of possible mismatch with long period variables, and a case having potential issues with the data.
           \modif{The colors of the models are the same as in Fig.~\ref{Fig:flcsSignificantComponents}.}
           }
\label{Fig:flcsInstrinsicScatter}
\end{figure*}

\begin{figure}
  \centering
  \includegraphics[width=1.00\columnwidth]{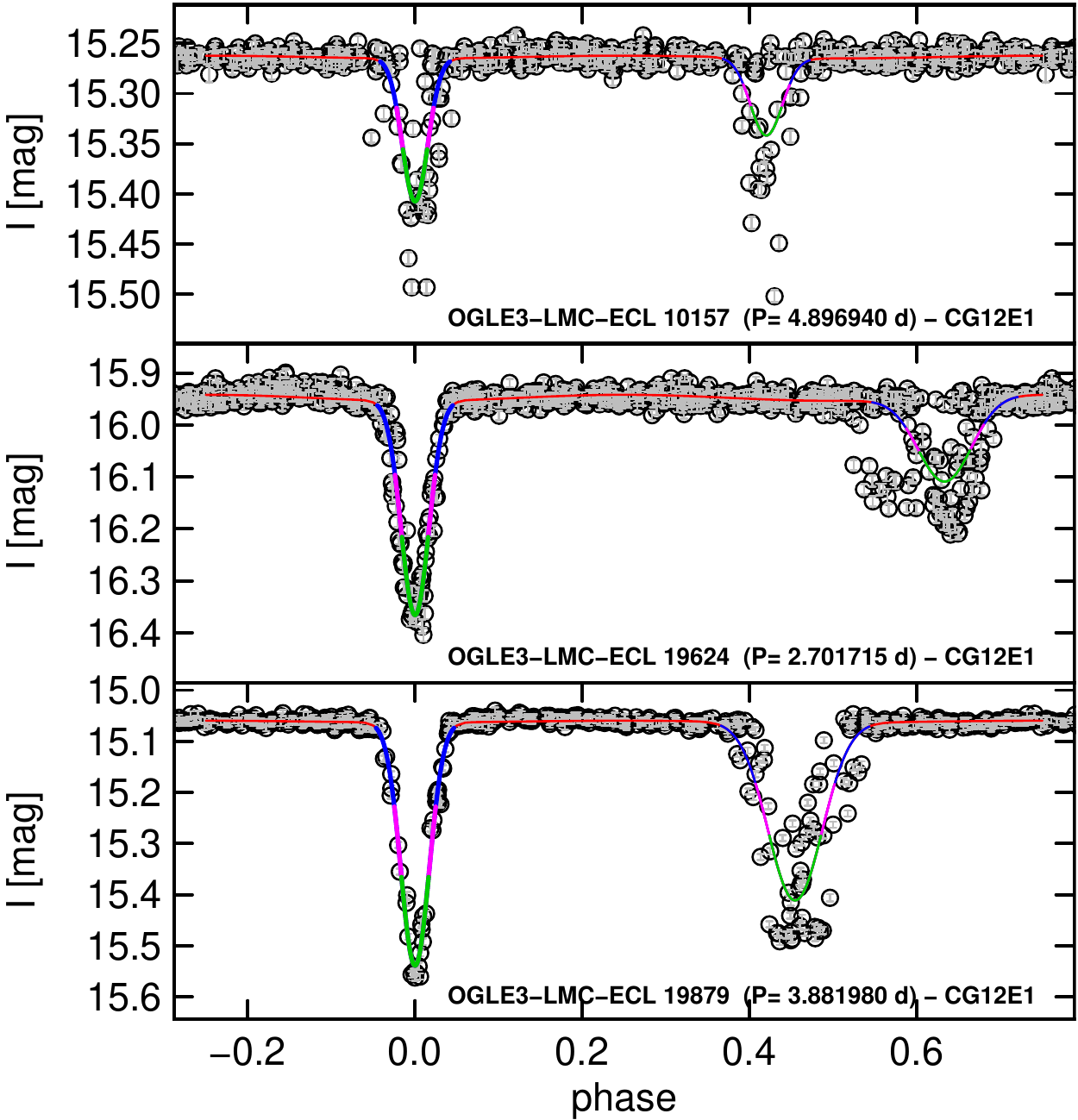}
  \caption{Examples of folded light curves with a strong scatter during the secondary eclipse, that leads to a large reduced $\chi^2$.
           \modif{The colors of the models are the same as in Fig.~\ref{Fig:flcsSignificantComponents}.}
           Their positions in the $\chi_\mathrm{reduced}^2$ versus $\Ab_\mathrm{resFLC}$ diagram are labelled in Fig.~\ref{Fig:reducedChi2VsAbbe_outliers}.
           }
\label{Fig:flcsPerturbedEclipse}
\end{figure}

\begin{figure}
  \centering
  \includegraphics[width=1.00\columnwidth]{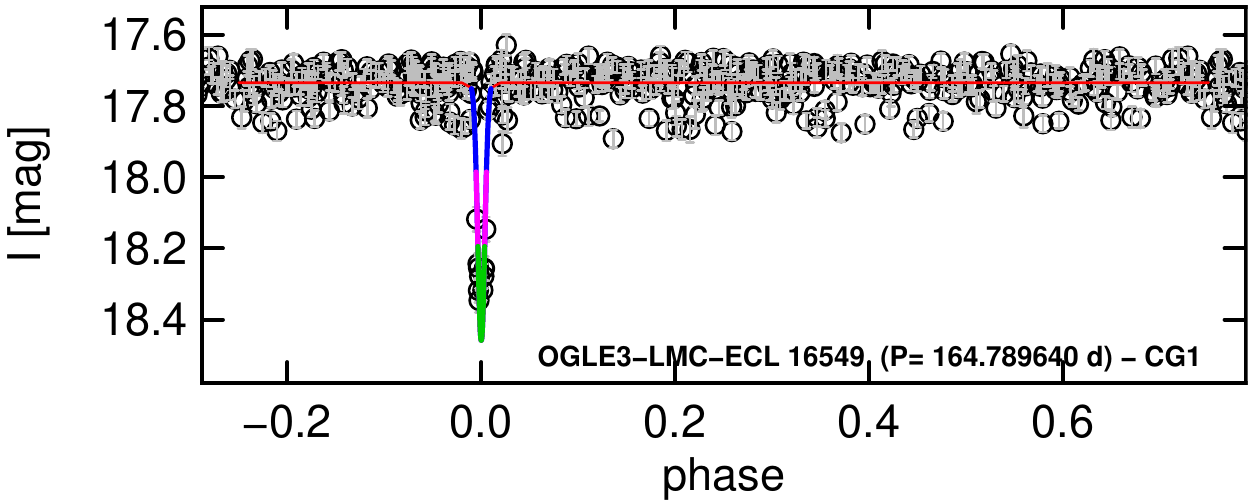}
  \caption{Light curve of quadruple system 16549, composed of two eclipsing binaries, folded with the period of the eclipsing binary with the longest period.
           \modif{The colors of the models are the same as in Fig.~\ref{Fig:flcsSignificantComponents}.}
           }
\label{Fig:flcsMultipleSystem}
\end{figure}

The $\chi_\mathrm{reduced}^2$ versus $\Ab_\mathrm{resFLC}$ diagram introduced in Sect.~\ref{Sect:fitQuality} offers a second interesting tool to identify outlying cases of EB LCs.
A value larger than 0.8 for $\Ab_\mathrm{resFLC}$ indicates a reasonable fit of the geometry of the initial FLC by the two-Gaussian model.
The resulting reduced $\chi^2$ should then be small.
For a fraction of the LCs, however, $\chi_\mathrm{reduced}^2$ is still large, as has been seen in Fig.~\ref{Fig:reducedChi2VsAbbe_2G}.
A residual scatter is thus present with an amplitude larger than expected from the measurement uncertainties.
Here, we check the status of those stars through a visual check of their LCs.

We select all stars that have $\Ab_\mathrm{resFLC}>0.8$ and $\chi_\mathrm{reduced}^2 > 10$, and identify example cases illustrating various potential origins for the higher-than-expected scatter in the residual FLC.
The sources chosen as examples are highlighted in the $\chi_\mathrm{reduced}^2$ versus $\Ab_\mathrm{resFLC}$ diagram shown in Fig.~\ref{Fig:reducedChi2VsAbbe_outliers}, and their LCs are shown in Figs.~\ref{Fig:flcsInstrinsicScatter} and \ref{Fig:flcsMultipleSystem}.
The following cases are identified.

\paragraph{Intrinsic periodic variability} --
If one or both stars are intrinsically variable, a residual scatter is naturally expected in the FLC.
This may be the case for source 16745 shown in Fig.~\ref{Fig:flcsRegionCIntrinsicScatter}, which has a large $\chi_\mathrm{reduced}^2$ (see Fig.~\ref{Fig:reducedChi2VsAbbe_outliers}).
Two other cases, sources 22288 and 2547, are shown in the top panels of Fig.~\ref{Fig:flcsInstrinsicScatter}, with intrinsic variability time scales long enough to be visible in their LCs.

\paragraph{Intrinsic non-periodic variability} --
An aperiodic variability can originate, e.g., from flares, outbursts, or irregular variability.
We visually identified several EBs presenting flares in the selected region of the $\chi_\mathrm{reduced}^2$ versus $\Ab_\mathrm{resFLC}$ diagram.
An example is shown in Fig.~\ref{Fig:flcsInstrinsicScatter} with source 7551.
It is characterized by bright outlying measurements apparently randomly distributed in the FLC.
The LC reveals three flares with time scales of the order of 100 days.
The source is blue ($V-I = -0.206$~mag) and has $\chi_\mathrm{reduced}^2=14.5$.

Two examples showing an outburst are given in Fig.~\ref{Fig:flcsInstrinsicScatter} with sources 18577 and 2594.
The LC shapes of these particular cases resemble that of microlensing events.
However, being blue ($V-I = -0.153$ and $-0.047$~mag, respectively), they may be blue bumpers \citep{CookAlcockAllsman_etal95,WyrzykowskiKozlowskiSkowron_etal11}, as also suggested by \cite{GraczykSoszynskiPoleski11} for source 2594.

Finally, an example of a source with irregular intrinsic variability is shown in Fig.~\ref{Fig:flcsInstrinsicScatter} with source 9699.
It shows irregular brightening and fading, on time scales of tens of days for the brightenings and hundred of days for the fadings.
The EB most probably hosts a Be star with a moderately blue color of $V-I=0.15$.
It has $\chi_\mathrm{reduced}^2 = 21.9$

\paragraph{Apsidal motions} --
Apsidal motion systems result from the rotation of the line of apsides, which is a consequence of non-axial distribution of component mass, leading to torque exerted on the Runge-Lenz vector.
It can be effectively modeled by a linear rate of change of the argument periastron, which manifests itself as an eclipse timing variation that causes both eclipses to excurse with respect to one another from their initial position.
Thus, both eclipses witness phase shifts, leading to a measurement of an anomalous orbital period when the phase of the system is defined with respect to superior conjunction.
We found almost twenty sources showing in-eclipse scatter of the measurements with an amplitude larger than expected from the out-of-eclipse scatter.
Three examples are shown in Fig.~\ref{Fig:flcsPerturbedEclipse} with sources 10157, 19624 and 19879.

\paragraph{Multiple systems} --
Multiple systems can reveal themselves through the presence of several periods in the LC for specific orbit configurations with respect to the line of sight.
Source 16549 shown in Fig.~\ref{Fig:flcsMultipleSystem} is an example of a hierarchical, gravitationally bound system that imprints its signature in the LC.
The four-body system is composed of two EB components, one with a long period reported in the OGLE-III catalog to be of 164.79~d, and a second one with a short period of 0.818033~d.
The LC of the system folded on the long period, shown in Fig.~\ref{Fig:flcsPerturbedEclipse}, clearly shows a narrow eclipse caused by the long-period binary component.
The period could actually be double this value, which would then reveal the presence of two eclipses in the FLC.
An analysis of the residual LC performed by \cite[][see in particular their Fig.~11]{GraczykSoszynskiPoleski11} reveals the presence of the additional short-period, \textit{EB}-type, contact system.
The contact binary introduces a scatter of $\sim$0.15~mag in the residual LC of the long-period system (which has a primary depth of 0.68~mag), that translates to $\chi_\mathrm{reduced}^2=4.1$ using our two-Gaussian model for the long-period system.

\paragraph{Disks} --
The presence of disks around one or both stars in a binary system can reveal itself in the LC geometry in and/or around the eclipses.
A nice example is given by source 17782 displayed in Fig.~\ref{Fig:flcsRegionCIntrinsicScatter}.
The source has been discussed by \cite{GraczykSoszynskiPoleski11} who conclude on the presence of a disk that contributes to a wide plateau in the primary eclipse superposed on a narrower stellar eclipse, with a morphology of the disk-induced eclipse that changes with time (see their Fig.~13).
The source is easily identified as an outlier in the $\chi_\mathrm{reduced}^2$ versus $\Ab_\mathrm{resFLC}$ diagram of Fig.~\ref{Fig:reducedChi2VsAbbe_outliers}, with $\chi_\mathrm{reduced}^2 = 41.8$.

\paragraph{Misclassification} --
A large $\chi_\mathrm{reduced}^2$ value can also result from a misclassification of the source.
Sources 957 and 7435, for example, shown in Fig.~\ref{Fig:flcsInstrinsicScatter}, have variable light variation amplitudes with time and may be long period variables (LPVs) instead of EBs.
This would be consistent with the red color of the two sources) ($V-I = 1.87$ and 2.21~mag, respectively) and their long periods of variability.
We note that the period of source 957 would then be $\sim$207~d if it was a LPV, i.e. half the quoted value of $\sim$414~d in the OGLE-III catalog.
In addition, a variability on time scales of several dozens of days is visible in the FLC and LC of this source.

\paragraph{Potential data reduction issues} --
Finally, the $\chi_\mathrm{reduced}^2$ versus $\Ab_\mathrm{resFLC}$ diagram can also serve as a diagnostic tool for data reduction quality.
Problems in data reduction will lead to artificially increased scatter in the residual LC.
The source will then appear as an outlier in the diagram, like the other sources analyzed above.
An example is shown in Fig.~\ref{Fig:flcsInstrinsicScatter} with source 11591, which displays a doubling of the 
LC towards the end of the OGLE-III survey.
Few such cases
 have been identified from our visual inspection of the selected region of outliers in the $\chi_\mathrm{reduced}^2$ versus $\Ab_\mathrm{resFLC}$ diagram.

\subsection{Statistical analysis}
\label{Sect:discussion_statisticalAnalysis}

We illustrate in this section the usage of the two-Gaussian models by analyzing the projected orbital eccentricities and eclipse widths of all models containing two significant eclipse candidates.
We filter the initial data set of all OGLE-III EBs of the LMC in several steps.
We first select all EBs for which the two-Gaussian model successfully describes the geometry of the FLC (i.e. $\Ab_\mathrm{resFLC}>0.8$, see Sect.~\ref{Sect:discussion_outliers_Abbe}) and which have a scatter in the residual LC (see Sect.~\ref{Sect:discussion_outliers_chi2}) smaller than $\chi_\mathrm{reduced}^2 = 5$.
This set contains 92\% of the initial OGLE-III catalog of LMC EBs.
We then take all EBs that are modeled with two Gaussians \modif{(i.e. for which the CG12, CG12E1 or CG12E2 has the largest BIC)}.
This represents 85\% of the previous set.
Finally, we restrict to models having significant eclipse candidates.
We use the significance criterion based on the $\Delta_\mathrm{Ecl1} \mathrm{BIC}$ and $\Delta_\mathrm{Ecl2} \mathrm{BIC}$ quantities introduced in Sect.~\ref{Sect:componentsSignificance}.
For the illustrative purposes of this section, we retain only models which have $\Delta_\mathrm{Ecl1} \mathrm{BIC}>50$ and  $\Delta_\mathrm{Ecl2} \mathrm{BIC}>50$ (see histograms of those quantities in Fig.~\ref{Fig:histoComponentsReliability}).
This represents \modif{77}\% of the previous set of models containing two Gaussians.
In total, our final set of EBs containing two significant eclipse candidates contains \modif{15681} sources.


\begin{figure}
  \centering
  \includegraphics[width=1.00\columnwidth]{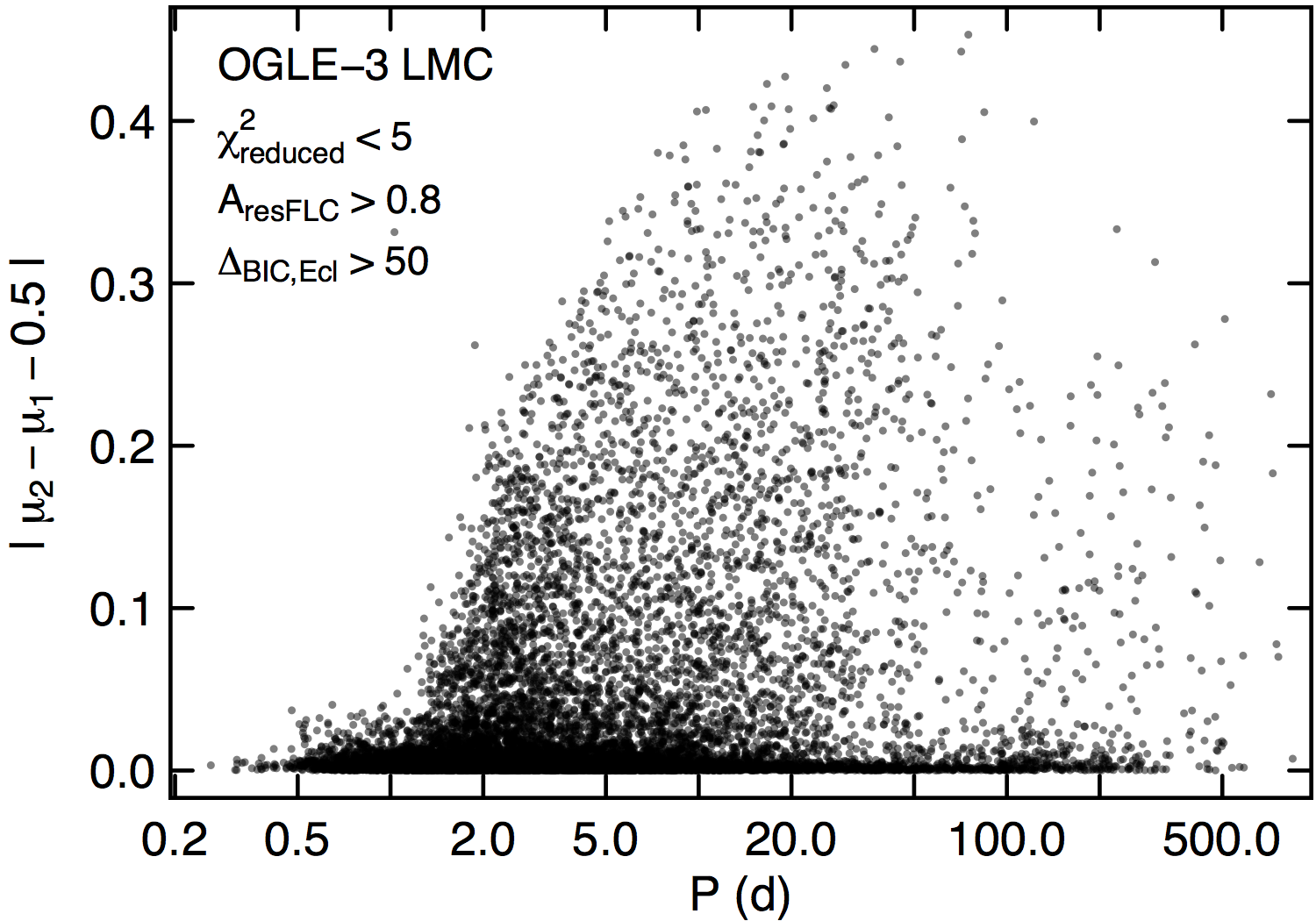}
  \includegraphics[width=1.00\columnwidth]{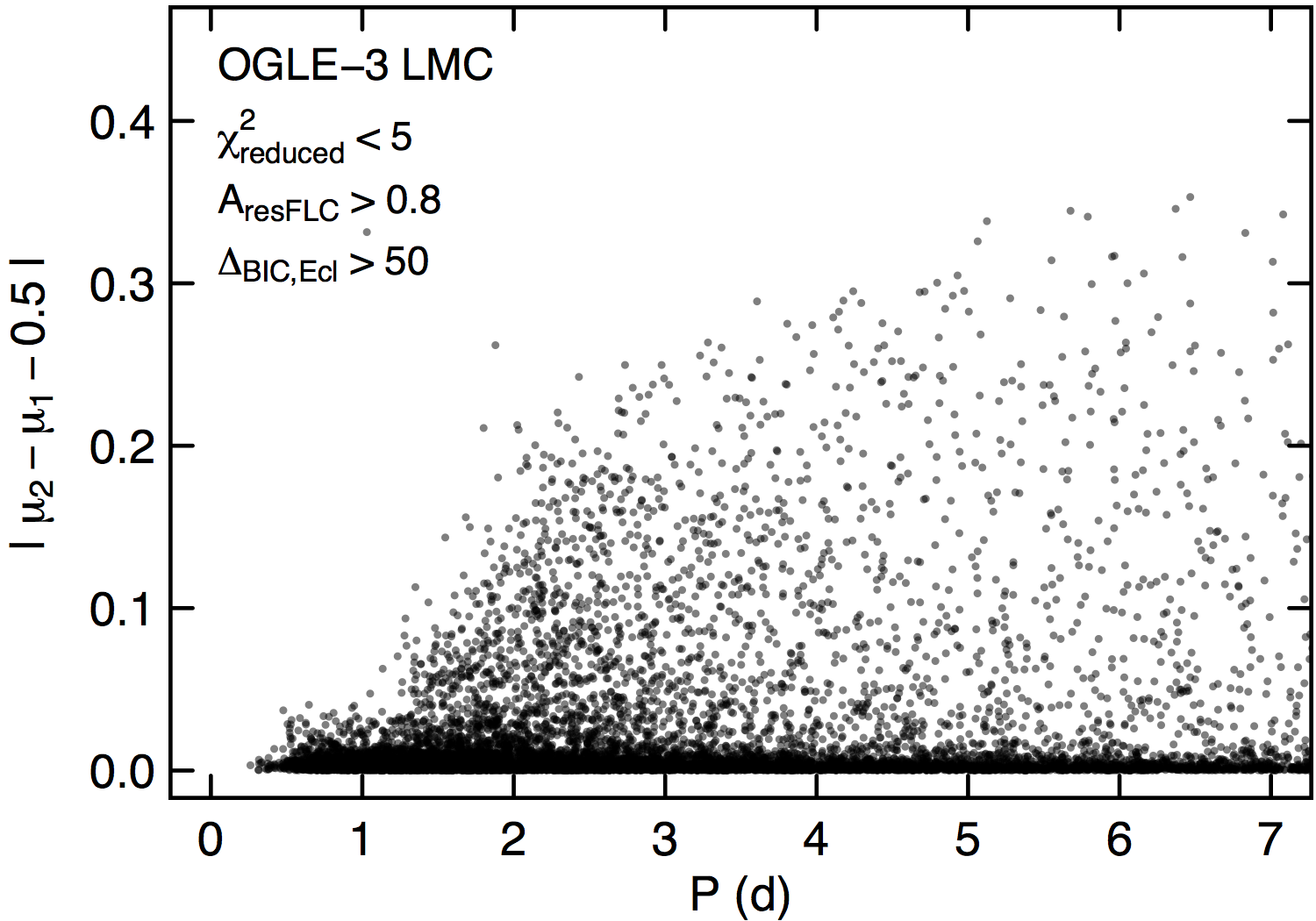}
  \caption{\textbf{Top panel:} Projected eccentricity, measured by the deviation $|\mu_2 - \mu_1 - 0.5|$ of the eclipse separation in phase with respect to 0.5, versus orbital period for all sources with $\Ab_\mathrm{resFLC}>0.8$ and $\chi_\mathrm{reduced}^2 < 5$ that have two significant eclipse candidates in their two-gaussian model (with the criterion $\Delta_\mathrm{Ecl1} \mathrm{BIC}>50$ and  $\Delta_\mathrm{Ecl2} \mathrm{BIC}>50$).
           \textbf{Bottom panel:} Same as the top figure, but zoomed on short orbital periods on a linear scale.
           }
\label{Fig:periodEccentricity}
\end{figure}

We take the deviation $|\mu_2 - \mu_1 - 0.5|$ of the eclipse separation in phase with respect to 0.5 as a proxy for the projected eccentricity.
This quantity is plotted versus the orbital period in the top panel of Fig.~\ref{Fig:periodEccentricity}.
The circularization of the orbit as the period shrinks is well visible in the figure.
A zoom at short periods is shown in the bottom panel of that figure.
The number of eccentric binaries decreases drastically at periods below 2 days, and all binary systems are found to be circular for periods shorter than $\sim$1.2~d.


\begin{figure}
  \centering
  \includegraphics[width=1.00\columnwidth]{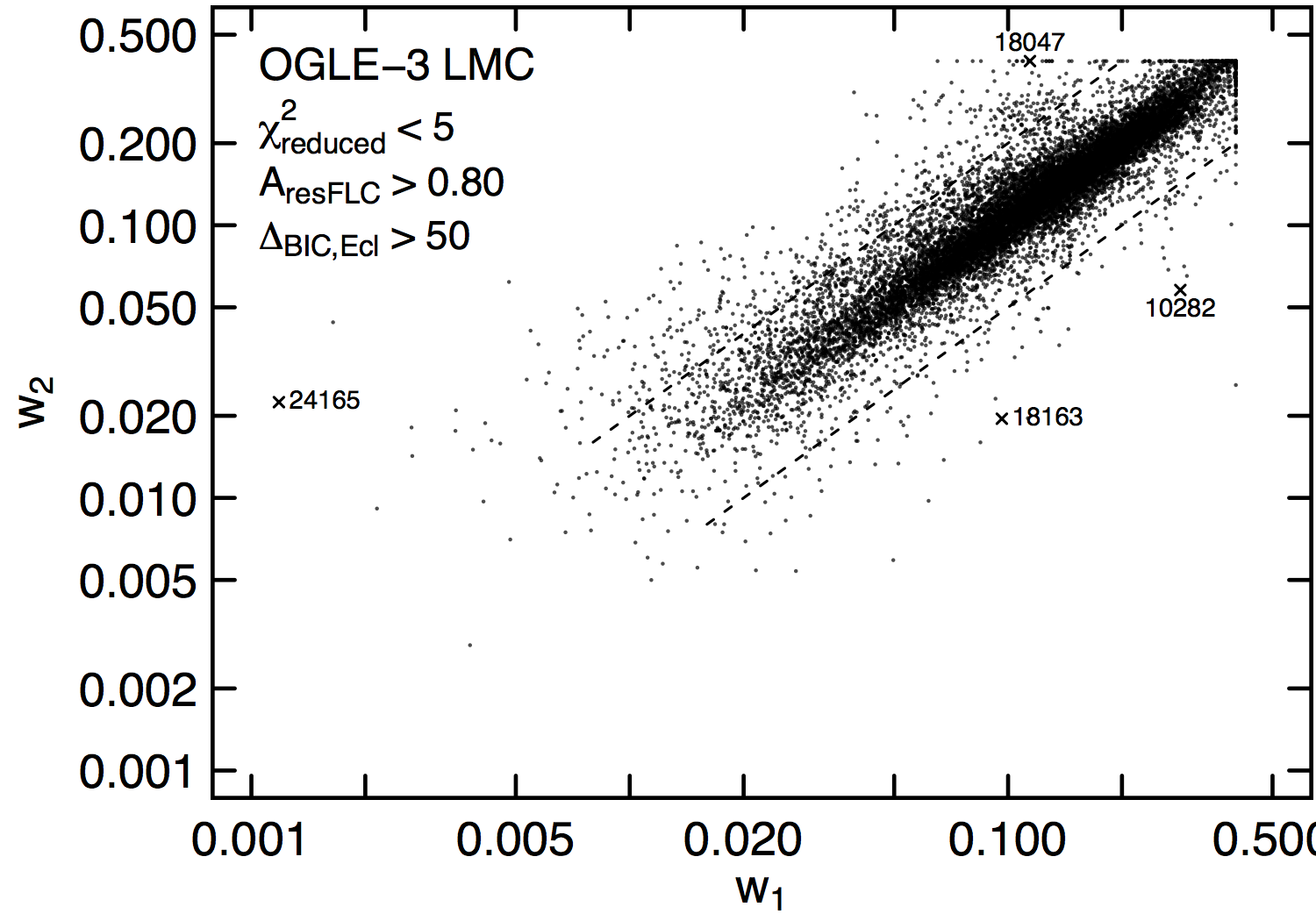}
  \caption{Same as Fig.~\ref{Fig:periodEccentricity}, but for the width (in phase) of primary eclipse versus the width (in phase) of secondary eclipse.
           The dashed lines locate ratios of primary over secondary eclipse widths equal to 0.5 and 2.
           Labelled sources are identified with cross marks and have their folded light curves shown in Fig.~\ref{Fig:flcsWidth1Width2}.
           }
\label{Fig:Width1Width2}
\end{figure}

\begin{figure}
  \centering
  \includegraphics[width=1.00\columnwidth]{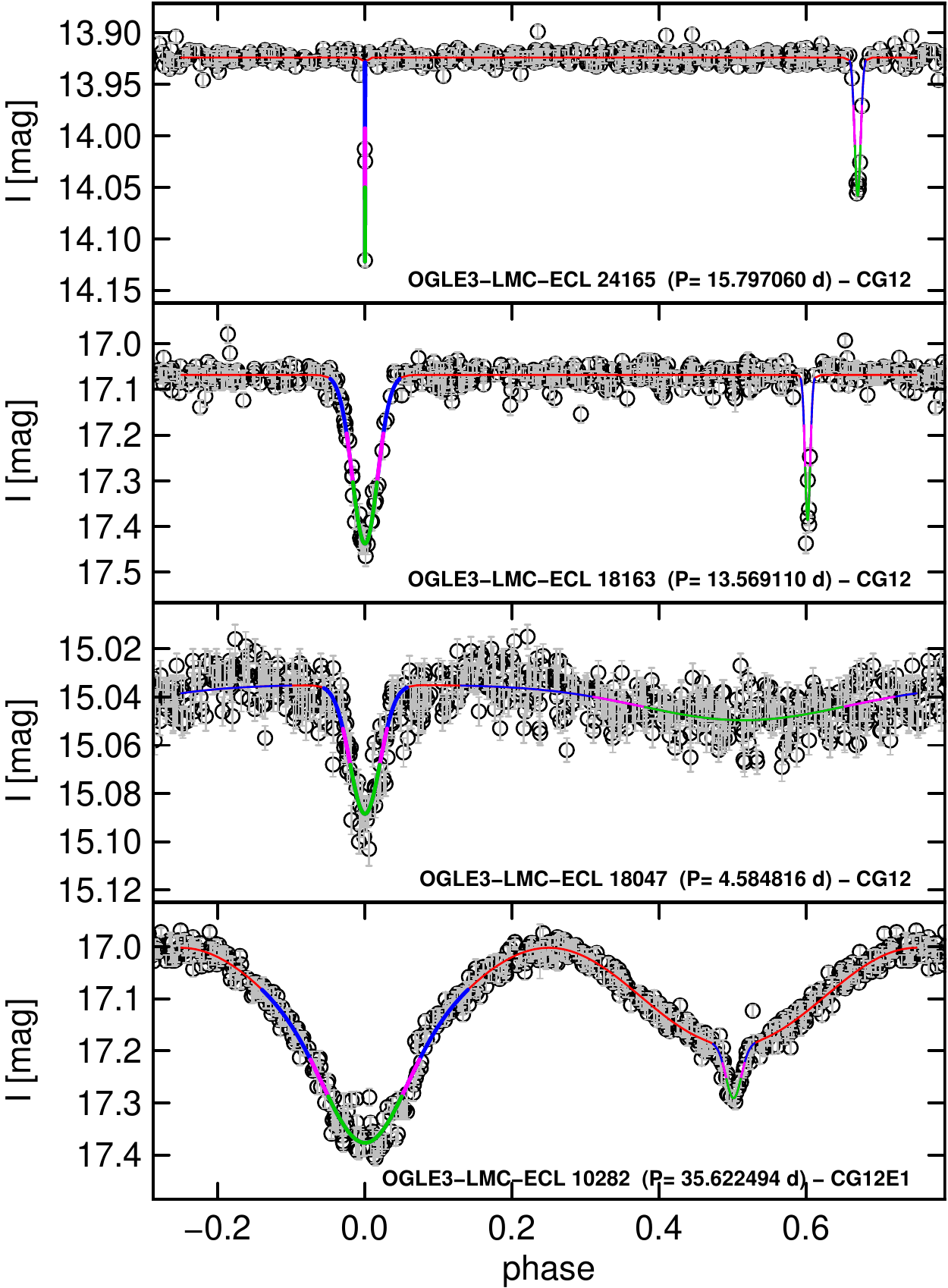}
  \caption{Folded light curves of various cases labeled in Fig.~\ref{Fig:Width1Width2}.
           \modif{The colors of the models are the same as in Fig.~\ref{Fig:flcsSignificantComponents}.}
           }
\label{Fig:flcsWidth1Width2}
\end{figure}

\begin{figure}
  \centering
  \includegraphics[width=1.00\columnwidth]{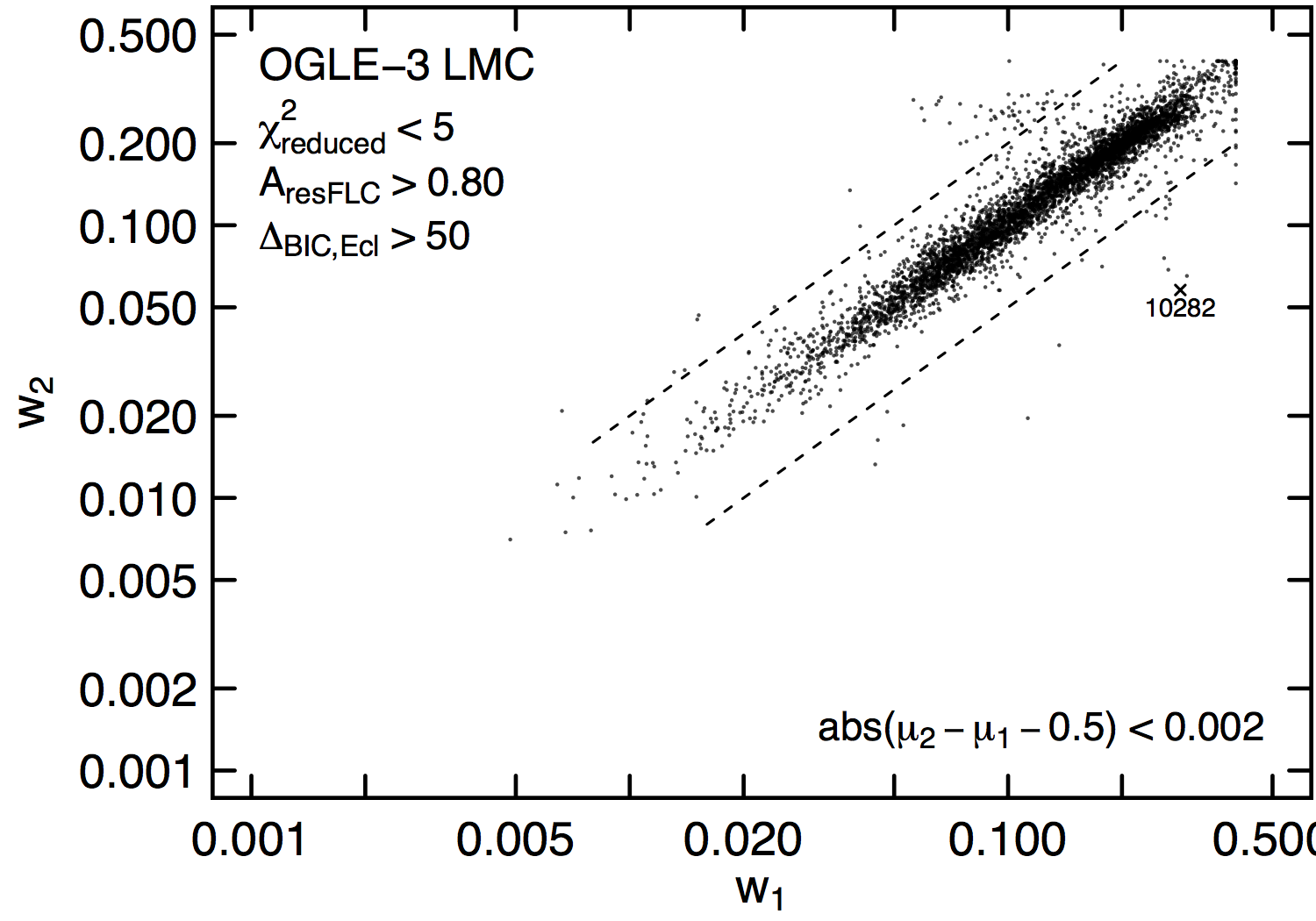}
  \includegraphics[width=1.00\columnwidth]{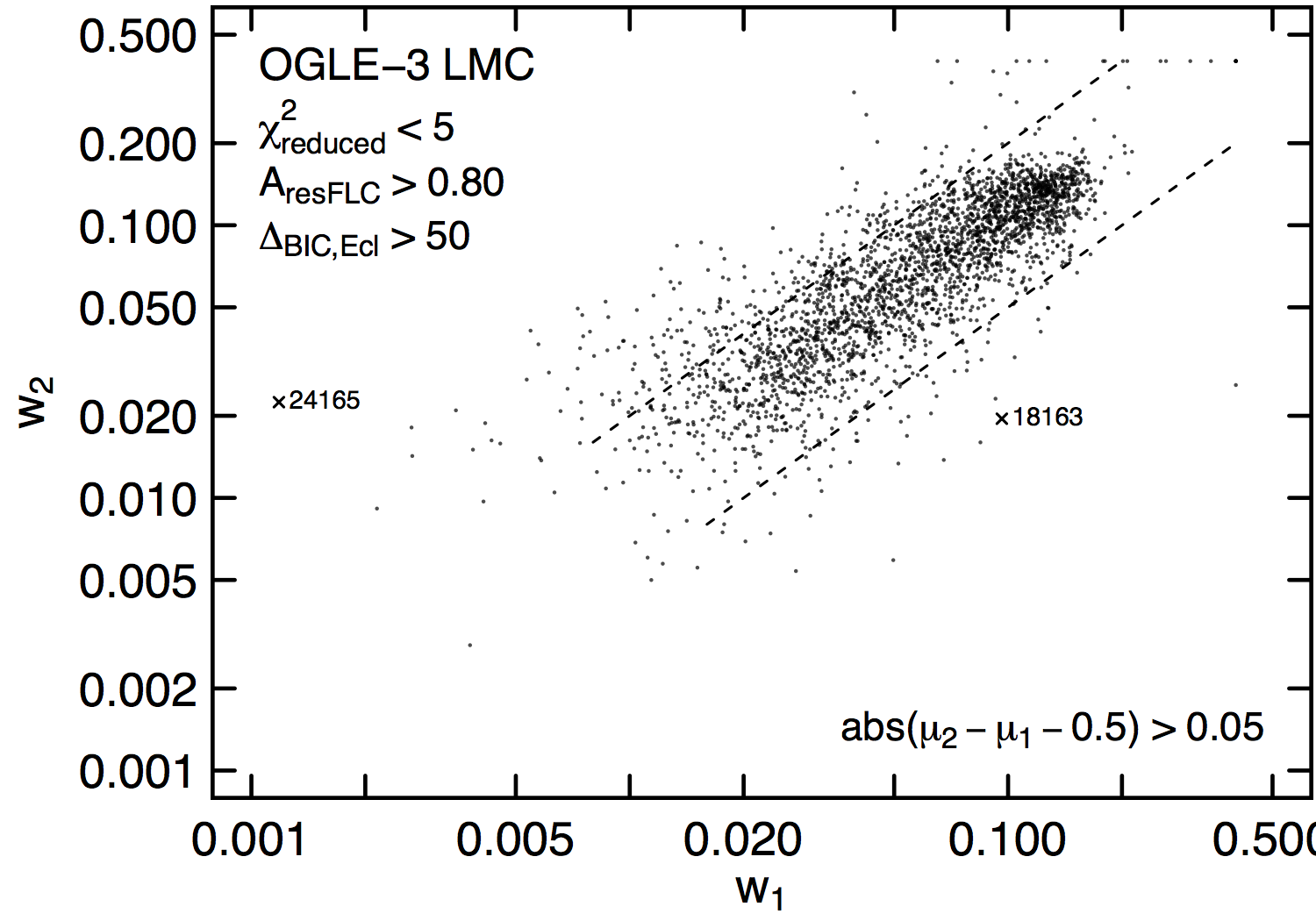}
  \caption{Same as Fig.~\ref{Fig:Width1Width2}, but restricted to systems with $|\mu_2 - \mu_1 - 0.5| < 0.01$ (top panel) or $|\mu_2 - \mu_1 - 0.5| > 0.05$ (bottom panel).
           }
\label{Fig:Width1Width2_restrictedEccentricities}
\end{figure}

Figure~\ref{Fig:Width1Width2} shows the widths $w_1$ and $w_2$ of the eclipses (see Eq.~\ref{Eq:eclipseWidth}).
As expected, the majority of EBs have primary and secondary eclipse widths that are within a factor of two of each other (i.e. within the area delimited by the dashed lines in Fig.~\ref{Fig:Width1Width2}) .
Some examples that deviate from this rule are shown in Fig.~\ref{Fig:flcsWidth1Width2}.
The two cases in the top two panels have very narrow eclipses, and are on an elliptic orbit.
The third case from top has an out-of-primary-eclipse geometry that is modeled by a wide second Gaussian, possibly representing an eccentric system with reflection.
The last case in the bottom panel of Fig.~\ref{Fig:flcsWidth1Width2} shows an ellipsoidal variability with large amplitude, possibly representing eccentric system with ellipsoidal variability.

The difference of eclipse width distributions between eccentric and circular systems is illustrated in Fig.~\ref{Fig:Width1Width2_restrictedEccentricities}.
The top panel of the figure shows systems with $|\mu_2 - \mu_1 - 0.5| < 0.002$, which favors\footnote{Eccentric systems with the longitude of periastron close to $\pm \pi/2$ will also satisfy $|\mu_2 - \mu_1 - 0.5| < 0.002$)} systems in circular orbits.
The bottom panel shows all systems with $|\mu_2 - \mu_1 - 0.5| > 0.05$, which selects eccentric systems.

%

\section{Conclusions}
\label{Sect:conclusions}

This work has shown the potential of Gaussian and cosine functions to model the geometry of EB LCs resulting from eclipse and ellipsoidal-like variability, respectively.
Using the two-Gaussian model parameters, we were successful in achieving our two goals, i.e. to identify outliers in a large set of EBs, and to provide a data base for the study of EB parameters on a statistical ground.

Key to these achievements are two diagrams introduced in Sect.~\ref{Sect:fitQuality}.
The first is the $\Ab_\mathrm{resFLC}$ versus $\Ab_\mathrm{FLC}$ diagram (Fig.~\ref{Fig:AbbeVsAbbe_2G}) that enables to identify outliers in terms of deviation of the FLC geometry from what can be modeled with a combination of Gaussian and cosine functions.
The second diagram is the $\chi_\mathrm{reduced}^2$ versus $\Ab_\mathrm{resFLC}$ diagram (Fig.~\ref{Fig:reducedChi2VsAbbe_2G}) to identify EBs that contain additional intrinsic variability other than that resulting from the binary nature of the source.
Those two diagrams have been exploited in Sect.~\ref{Sect:discussion_outliers} to identify potentially interesting binary systems.

Section~\ref{Sect:discussion_statisticalAnalysis} has then briefly illustrated how the two-Gaussian model results can be used to study the properties of EBs on a statistical ground.
An inevitable challenge of automated procedures is to minimize as much as possible the contamination of statistical conclusions by the presence of non-physical components in the models.
We presented in Sect.~\ref{Sect:componentsSignificance} a method based on BIC analysis to estimate the significance of each component in the two-Gaussian models.
In particular, the significances of primary ($\Delta_\mathrm{Ecl1} \mathrm{BIC}$) and secondary ($\Delta_\mathrm{Ecl2} \mathrm{BIC}$) eclipse candidates can be used to filter out models that have a high probability to contain spurious eclipses.

The results of our two-Gaussian models for the OGLE-III EBs of the LMC are available in electronic format.
A description of its content is given in Table~\ref{Tab:results} in the Appendix.

This work constitutes a basis for the establishment of an automated pipeline to process \modif{{\Gaia} LCs}.
\modif{
{\Gaia} LCs will have, on the mean, about 70 measurements on a 5-year mission.
The efficiency of the two-Gaussian model to characterize the LCs of the EBs seen by {\Gaia} has been addressed by \cite{KochoskaMowlaviPrsa_etal17}.
In that study, the two-Gaussian method has been applied to both the original Kepler LCs and to the Kepler set of LCs resampled with {\Gaia} cadence using the {\Gaia} scanning law at the sky position of the Kepler EBs, and considering a 5-year time span for the {\Gaia} mission.
The study reveals that 2/3 of the Kepler EBs are detectable by {\Gaia} due to the presence of a sufficient number of observations in the eclipses.
The study further shows that, when this is the case, the two-Gaussian method is successful in characterizing the LC geometry of the EBs.
\cite{KochoskaMowlaviPrsa_etal17} further propose a classification scheme of the detectable sources based on the morphological type indicative of the light curve.
}

Several improvements \modif{to the two-Gaussian} model are foreseen for the next steps.
They comprise the inclusion of an additional component in the models to describe reflection.
We also pursue our exploratory works of automated classification techniques initiated with the works of \cite{KochoskaMowlaviPrsa_etal17} and \cite{SuvegesBarblanLecoeurTaibi_etal17}.
On the path to these realizations, the various procedures will be applied and tested on existing data from surveys such as Hipparcos and the recently-released OGLE-IV, as well as on {\Gaia}-simulated data.

%

\bibliographystyle{aa}
\bibliography{bibTex}

\begin{appendix}

\section{Table description of the two-Gaussian model results}
\label{Appendix:tableDescription}

\begin{table*}
\centering
\caption{Two-Gaussian model attributes published in electronic format.}
\begin{tabular}{l l c l}
\hline
Attribute        & Symbol                      & Unit & Description\\
                 & in text                     &      & \\
\hline
sourceId         &                             & -    & OGLE-III eclipsing binary source ID in the LMC\\
period           & $P$                         & day  & Orbital period\\
model            &                             & -    & Adopted two-Gaussian model (see Table~\ref{Tab:models})\\
numParams        & $p$                         & -    & Number of parameters in the two-Gaussian model\\
primaryEpoch     &                             & day  & Epoch of primary minimum (HJD-2450000)\\
cst              & $C$                         & mag  & Value of the constant in the two-Gaussian model\\
cstErr           &                             & mag  & Uncertainty on C\\
mu1              & $\mu_1$                     & -    & Phase of primary eclipse minimum\\
mu1Err           & $\mu_\mathrm{1,err}$        & -    & Uncertainty on mu1\\
sigma1           & $\sigma_1$                  & -    & Gaussian width, in phase, of the primary eclipse\\
sigma1Err        & $\sigma_\mathrm{1,err}$     & -    & Uncertainty on sigma1\\
d1               & $d_1$                       & mag  & Gaussian depth of the primary eclipse\\
d1Err            & $d_\mathrm{1,err}$          & -    & Uncertainty on d1\\
mu2              & $\mu_2$                     & -    & Phase of secondary minimum\\
mu2Err           & $\mu_\mathrm{2,err}$        & -    & Uncertainty on mu2\\
sigma2           & $\sigma_2$                  & -    & Gaussian width, in phase, of the secondary eclipse\\
sigma2Err        & $\sigma_\mathrm{2,err}$     & -    & Uncertainty on sigma2\\
d2               & $d_2$                       & mag  & Gaussian depth of the secondary eclipse\\
d2Err            & $d_\mathrm{2,err}$          & -    &  Uncertainty on d2\\
muForCosHalfP    & $\varphi_\mathrm{0,ell}$    & -    & Phase of cosine function for the ellipsoidal component\\
aCosHalfP        &$\frac{1}{2}A_\mathrm{0,ell}$& mag  & Amplitude of cosine function for the ellipsoidal component\\
aCosHalfPErr     &                             & mag  & Uncertainty on aCosHalfP\\
width1           & $w_1$                       & -    & Primary eclipse duration in phase\\
depth1           & $d'_1$                      & mag  & Primary eclipse depth\\
width2           & $w_2$                       & -    & Secondary eclipse duration in phase\\
depth2           & $d'_2$                      & mag  & Secondary eclipse depth\\
maxPhaseGap      &                             & -    & Largest phase gap in folded light curve\\
phaseClumpiness  &                             & -    & Phase clumpiness\\
phaseCoverageEcl1&                             & -    & Phase coverage of primary eclipse\\
phaseCoverageEcl2&                             & -    & Phase coverage of secondary eclipse\\
significance\_ecl1& $\Delta_\mathrm{ecl1} \mathrm{BIC}$ & - & Significance of primary eclipse\\
significance\_ecl2& $\Delta_\mathrm{ecl2} \mathrm{BIC}$ & - & Significance of secondary eclipse\\
significance\_ell & $\Delta_\mathrm{ell} \mathrm{BIC}$ & - & Significance of ellipsoidal-like variability\\
ecl1\_dOverMeanMagError & $d_1/\bar{\varepsilon}_{\mathrm{ecl},1}$ & - & Gaussian depth over mean measurement uncertainty for primary eclipse\\
ecl2\_dOverMeanMagError & $d_2/\bar{\varepsilon}_{\mathrm{ecl},2}$ & - & Gaussian depth over mean measurement uncertainty for secondary eclipse\\
abbeFlc          & $\Ab_\mathrm{FLC}$          & -    & Abbe value of the folded light curve\\
abbeFlcResidual  & $\Ab_\mathrm{resFLC}$       & -    & Abbe value of the residual folded light curve\\
reducedChi2      & $\chi_\mathrm{reduced}^2$    & -    & Reduced $\chi^2$\\
bic\_CG12        & BIC$_\mathrm{CG12}$         & -    & Bayesian information criterion value of the CG12 model\\
bic\_CG12E1      & BIC$_\mathrm{CG12E1}$       & -    & Bayesian information criterion value of the CG12E1 model\\
bic\_CG12E2      & BIC$_\mathrm{CG12E2}$       & -    & Bayesian information criterion value of the CG12E2 model\\
bic\_CG1         & BIC$_\mathrm{CG1}$          & -    & Bayesian information criterion value of the CG1 model\\
bic\_CG1E1       & BIC$_\mathrm{CG1E1}$        & -    & Bayesian information criterion value of the CG1E1 model\\
bic\_CG2         & BIC$_\mathrm{CG2}$          & -    & Bayesian information criterion value of the CG2 model\\
bic\_CG2E2       & BIC$_\mathrm{CG2E2}$        & -    & Bayesian information criterion value of the CG2E2 model\\
bic\_CE          & BIC$_\mathrm{CE}$           & -    & Bayesian information criterion value of the CE model\\
bic\_C           & BIC$_\mathrm{C}$            & -    & Bayesian information criterion value of the C model\\
\hline
\end{tabular}
\label{Tab:results}
\end{table*}

Table~\ref{Tab:results} summarizes the content of the electronic table giving the two-Gaussian models for all the OGLE-III LMC EBs.
In the electronic version, a 'NA' is published when a quantity is not applicable for a given source, for example for the parameters of a secondary eclipse when the model contains only one Gaussian.
The table contains
\begin{itemize}
\item the source ID number and the orbital period given in the OGLE-III catalog for the LMC EBs;
\item the two-Gaussian model chosen by our automated procedure based on the BIC (see Sect.~\ref{Sect:models_selection});
\item the epoch of primary eclipse minimum and the two-Gaussian parameters defined in Eqs.~\ref{Eq:Gaussian} and \ref{Eq:gaussianFct}.
      The uncertainties associated to the parameters are taken from the covariance matrix returned by the non-linear fitting procedure;
\item the widths and depths of the eclipses, defined by Eqs.~\ref{Eq:eclipseWidth} and \ref{Eq:eclipseDepth}, respectively;
\item the maximum phase gap, the phase clumpiness, and the eclipse phase coverages defined in Sect.~\ref{Sect:phaseCoverage};
\item the significances $\Delta_\mathrm{component} \mathrm{BIC}$ of the two-Gaussian model components described in Sect.~\ref{Sect:componentsSignificance};
\item the depths of the eclipse candidates relative to the mean measurement uncertainties inside the eclipses, discussed in Sect.~\ref{Sect:modelsSignificance_eclDepth};
\item the Abbe and reduced $\chi^2$ values of the FLCs, introduced in Sect.~\ref{Sect:fitQuality} to evaluate the overall quality of the fits;
\item and the BIC values of all the two-Gaussian models evaluated for each EB.
      Models containing one Gaussian, named CG (CGE) in Table~\ref{Tab:models} when they do not (they do) contain an ellipsoidal-like variability, are divided in Table~\ref{Tab:results} into CG1 (CG1E1) and CG2 (CG2E2) depending on whether the unique Gaussian is centered on the first or second eclipse candidate identified in the initial value determination step of model parameters (see Sect.~\ref{Sect:procedure_initialValues}).
      These two initial parameter sets are tested in succession when evaluating for the best model.
      The distinction between between CG1 (CG1E1) and CG2 (CG2E2) models is also necessary for the computation of the eclipse significances $\Delta_\mathrm{ecl1} \mathrm{BIC}$ and $\Delta_\mathrm{ecl2} \mathrm{BIC}$ of models containing two Gaussians.
      If the initial value determination procedure of model parameters finds only one eclipse candidate, models CG2 and CG2E2 are non-existent.
      The BIC values of some models may also be inexistent if the non-linear procedure fails to converge.
\end{itemize}

\end{appendix}

\end{document}